\documentclass[twocolumn]{aastex701}

\graphicspath{{./}}

\newcommand{\sherpa}{Sherpa}

\usepackage{enumitem}

\received{17 April 2026}
\revised{8 May 2026}
\accepted{12 May 2026}
\submitjournal{The Astrophysical Journal}

\begin{document}

\title{CIAO: Chandra's Data Analysis System for X-Ray Astronomy and Beyond}

\correspondingauthor{Antonella Fruscione}
\author[orcid=0000-0002-6414-3954]{Antonella Fruscione}
\affiliation{Center for Astrophysics $|$ Harvard \& Smithsonian, 60 Garden St., Cambridge, MA 02138, USA}
\email[show]{afruscione@cfa.harvard.edu}

\author[orcid=0000-0002-7093-295X]{Jonathan McDowell}
\altaffiliation{Retired}
\affiliation{Center for Astrophysics $|$ Harvard \& Smithsonian, 60 Garden St., Cambridge, MA 02138, USA}
\email{jmcdowell@cfa.harvard.edu}
\affiliation{Durham University Space Research Center, Stockton Rd, Durham DH1 3LE, UK}

\author[orcid=0000-0003-4428-7835]{Douglas Burke}
\affiliation{Center for Astrophysics $|$ Harvard \& Smithsonian, 60 Garden St., Cambridge, MA 02138, USA}
\email{dburke@cfa.harvard.edu}

\author{Mark Cresitello Dittmar}
\affiliation{Center for Astrophysics $|$ Harvard \& Smithsonian, 60 Garden St., Cambridge, MA 02138, USA}
\email{mdittmar@cfa.harvard.edu}

\author[orcid=0000-0003-0494-743X]{Ian N. Evans}
\affiliation{Center for Astrophysics $|$ Harvard \& Smithsonian, 60 Garden St., Cambridge, MA 02138, USA}
\email{ievans@cfa.harvard.edu}

\author[orcid=0000-0003-3509-0870]{Janet D. Evans}
\affiliation{Center for Astrophysics $|$ Harvard \& Smithsonian, 60 Garden St., Cambridge, MA 02138, USA}
\email{jevans@cfa.harvard.edu}

\author[orcid=0009-0000-5187-673X]{Kenny Glotfelty}
\affiliation{Center for Astrophysics $|$ Harvard \& Smithsonian, 60 Garden St., Cambridge, MA 02138, USA}
\email{kglotfelty@cfa.harvard.edu}

\author[orcid=0000-0003-4243-2840]{Hans Moritz G\"unther}
\affiliation{MIT Kavli Institute for Astrophysics and Space Research, Massachusetts Institute of Technology, 70 Vassar St, Cambridge, MA 02139, USA}
\email{hmg@space.mit.edu}

\author[orcid=0000-0002-3860-6230]{David Huenemoerder}
\affiliation{MIT Kavli Institute for Astrophysics and Space Research, Massachusetts Institute of Technology, 70 Vassar St, Cambridge, MA 02139, USA}
\email{dph@space.mit.edu}

\author{William Joye}
\altaffiliation{Retired}
\affiliation{Center for Astrophysics $|$ Harvard \& Smithsonian, 60 Garden St., Cambridge, MA 02138, USA}
\email{wjoye@cfa.harvard.edu}

\author[orcid=0009-0006-5274-6439]{Nicholas P. Lee}
\affiliation{Center for Astrophysics $|$ Harvard \& Smithsonian, 60 Garden St., Cambridge, MA 02138, USA}
\email{nplee@cfa.harvard.edu}

\author{Warren McLaughlin}
\affiliation{Center for Astrophysics $|$ Harvard \& Smithsonian, 60 Garden St., Cambridge, MA 02138, USA}
\email{wmclaughlin@cfa.harvard.edu}

\author[orcid=0009-0009-9067-2030]{Joseph B. Miller}
\affiliation{Center for Astrophysics $|$ Harvard \& Smithsonian, 60 Garden St., Cambridge, MA 02138, USA}
\email{jmiller@cfa.harvard.edu}

\author[orcid=0000-0002-3310-1946]{Melania Nynka}
\affiliation{MIT Kavli Institute for Astrophysics and Space Research, Massachusetts Institute of Technology, 70 Vassar St, Cambridge, MA 02139, USA}
\email{nynka@space.mit.edu}

\author[orcid=0000-0002-7939-377X]{David A. Principe}
\affiliation{MIT Kavli Institute for Astrophysics and Space Research, Massachusetts Institute of Technology, 70 Vassar St, Cambridge, MA 02139, USA}
\email{principe@mit.edu}

\author[orcid=0000-0002-0905-7375]{Aneta Siemiginowska}
\affiliation{Center for Astrophysics $|$ Harvard \& Smithsonian, 60 Garden St., Cambridge, MA 02138, USA}
\email{asiemiginowska@cfa.harvard.edu}

\collaboration{30}{(Chandra X-Ray Center Science Data System Group and Data Systems Division)}

\begin{abstract}
The \textit{Chandra Interactive Analysis of Observations} (CIAO) software, developed by the Chandra X-ray Center, has been the data analysis package for the Chandra X-ray Observatory since its launch in 1999. Over nearly three decades, CIAO has grown from a small software suite into a widely used system for X-ray data analysis and beyond.

CIAO provides tools for calibration, spectral, imaging, and timing analysis, together with high-level scripts and the \sherpa\ modeling and fitting application. Its modular design and unified data model allow users to build flexible analysis workflows while maintaining consistency with the Chandra data processing pipeline. Visualization capabilities are provided through integration with SAOImageDS9 and Python-based tools, and simulation components such as ChaRT and MARX extend the analysis environment to include detailed modeling of instrumental effects.

In this paper we describe CIAO's design, evolution, and capabilities after 25 years of Chandra operations. We also describe its core architecture, scripting environment, modeling, visualization tools, simulation components, and testing infrastructure, as well as the documentation and user support system that have contributed to its widespread use. CIAO's continued development and broad adoption highlight its important role in X-ray astronomy and its usefulness in multiwavelength astrophysical research.
\end{abstract}

\keywords{\uat{Astronomy software}{1855} --- \uat{Open source software}{1866} --- \uat{Software documentation}{1869} --- \uat{X-ray astronomy}{1810}}

\section{Introduction}

The Chandra X-ray Observatory, a NASA flagship mission launched in 1999 \citep{weisskopf2002}, has been a cornerstone in high-energy astrophysics. Its exceptional angular resolution and sensitivity have enabled studies spanning a wide range of astrophysical environments, from planetary systems to distant quasars \citep{wilkes2019,slane2025}. X-ray data is inherently 4-dimensional, and with Chandra all four are commonly needed: right ascension (RA), declination (Dec), energy, and time. This necessitated a data analysis system that could handle higher dimensionality data. In addition, the analysis of X-ray data, characterized by low photon counts, complex instrumental effects, and the need for robust statistical methods, requires dedicated data analysis software \citep[e.g.,][]{handbook2011}.

The \textit{Chandra Interactive Analysis of Observations} (CIAO) software suite was developed at the Chandra X-Ray Center (CXC) to meet this need, as described in its original reference paper \citep{fruscione2006}. In this paper we describe CIAO in its present form (version 4.18.0)\footnote{\url{https://cxc.cfa.harvard.edu/ciao/}}, which incorporates significant changes since the first release, including the transition to a Python-based environment, the expansion of Sherpa---CIAO's modeling and fitting applications---and the development of high-level analysis scripts.

\begin{figure}[ht!]
\includegraphics[width=0.47\textwidth]{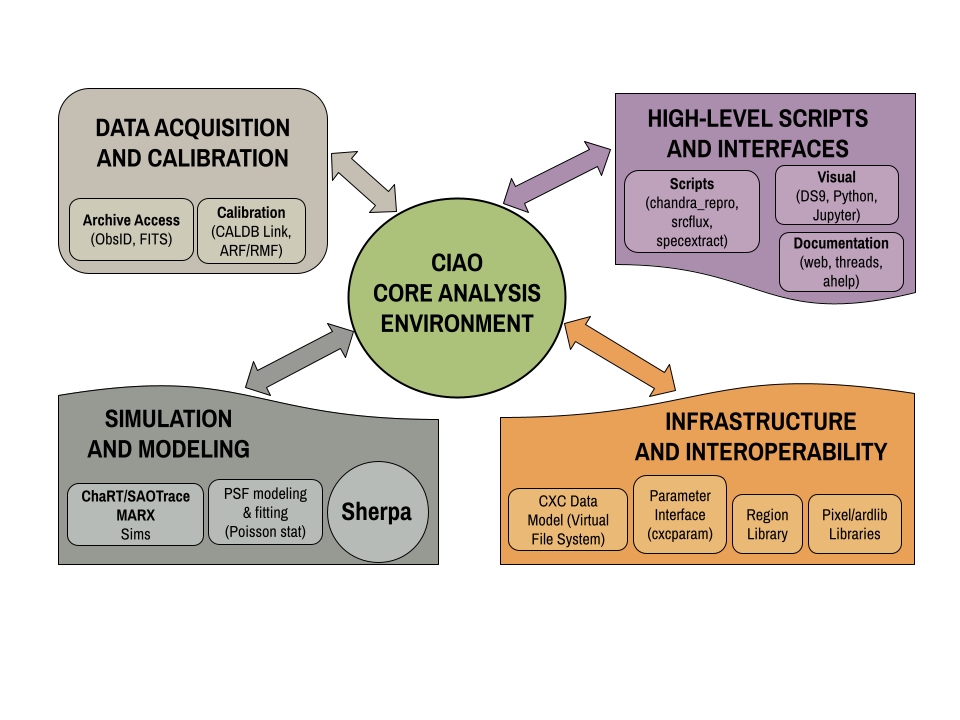}
\caption{High-level schematic of the CIAO data analysis system. This simplified diagram illustrates, in an approximate way, the bidirectional connections between the central analysis core and the specialized modules for data acquisition, modeling, infrastructure, and user workflows.}
\label{fig:bubble}
\end{figure}

In its current form, CIAO has become a flexible and extensible software suite. CIAO includes tools for calibration, spectral, imaging, and timing analysis---and combined analysis of these dimensions---of all Chandra instruments and gratings whether used for direct imaging or as grating readout detectors. CIAO's design combines data handling, processing, and modeling in a cohesive system, allowing users to carry out uniform analyses and to facilitate studies that combine data from multiple observatories (see e.g. \citealt{guver2022,hyeongHan2024,kovacs2025,wood2025}). CIAO and Sherpa adopt a different design approach compared to other widely used X-ray analysis tools. Packages such as the \textit{High Energy Astrophysics Software} (HEASoft\footnote{\url{https://heasarc.gsfc.nasa.gov/docs/software/heasoft/}}) and the \textit{XMM-Newton Science Analysis System} (SAS\footnote{\url{https://www.cosmos.esa.int/web/xmm-newton/sas}}; \citealt{gabriel2004}) are primarily organized around mission-specific pipelines, while the \textit{X-ray Spectral Fitting Package} (XSPEC; \citealt{arnaud1996}) is optimized for spectral analysis. CIAO is designed as an integrated environment with a strong emphasis on imaging, while also supporting spectral and timing analysis within a unified framework. Its abstract data model enables seamless transitions between data products, facilitating the combination of spatial, spectral, and temporal information. Sherpa extends this approach by providing a common statistical framework for modeling images, spectra, and time series. CIAO and Sherpa are therefore complementary to existing tools, and are particularly well suited to imaging-driven analyses and studies that combine multiple data types or instruments. A high-level overview of this integrated architecture, highlighting the bidirectional flow between the core analysis environment and specialized modules for calibration, simulations, and user interfaces, is shown in Figure~\ref{fig:bubble}.

\section{Design Principles and Evolution of CIAO}

The word ``\textit{ciao}'' derives from the old Venetian expression ``\textit{s'ciavo},'' meaning ``I am your servant,'' reflecting its role as a system designed to support Chandra users \citep{fruscione2000}. The initial CIAO concept was based on three key principles: modularity, user accessibility, and scientific rigor. CIAO was designed as a modular system in which users combine standalone tools for tasks such as event filtering, spectral extraction, and source detection. It also introduced ``analysis threads'' as a new paradigm: step-by-step, reproducible recipes to guide users through complete scientific analyses. Together with extensive documentation and tutorials, these features make the system accessible to a broad user base, from novice to expert \citep{burke2006,galle2003,galle2005}. From the beginning, CIAO incorporated advanced statistical methods to address novel challenges inherent in Chandra data \citep{siemiginowska1997}, with \sherpa\ providing modeling and fitting capabilities \citep{freeman2001,siemiginowska2024}.

CIAO Version 1.0 was released in October 1999, shortly after the launch of the Chandra X-Ray Observatory. CIAO, primarily implemented in C and C++, included approximately 30 tools and supported Solaris and early Red Hat Linux systems. Early CIAO development focused on data processing and calibration products underlying quantitative analysis, including tools to generate ancillary response files (ARFs), redistribution matrix files (RMFs), and point spread functions (PSFs) to account for Chandra's energy-dependent response. These capabilities included Chandra-specific features such as modeling of the line-spread function for high-resolution dispersed spectra and treatment of aspect dither, which averages responses over detector features. The formal framework for these response calculations is described by \cite{davis2001b}.

In December 2000, following a critical evaluation of interactive languages, S-Lang\footnote{\url{https://www.jedsoft.org/slang/}} was introduced in CIAO 2.0 for advanced scripting, offering an order-of-magnitude performance improvement over alternatives available at the time \citep{noble2008}.

A major transition occurred with CIAO 4.0 (2008), which introduced a redesigned architecture and replaced S-Lang with Python\footnote{\url{https://www.python.org/}} as the primary user-facing language. This shift aligned CIAO with developments in scientific computing allowing integration with libraries such as NumPy and Matplotlib \citep{deponte2008,refsdal2009,doe2009,galle2011}.

\begin{figure}[ht!]
\includegraphics[width=0.47\textwidth]{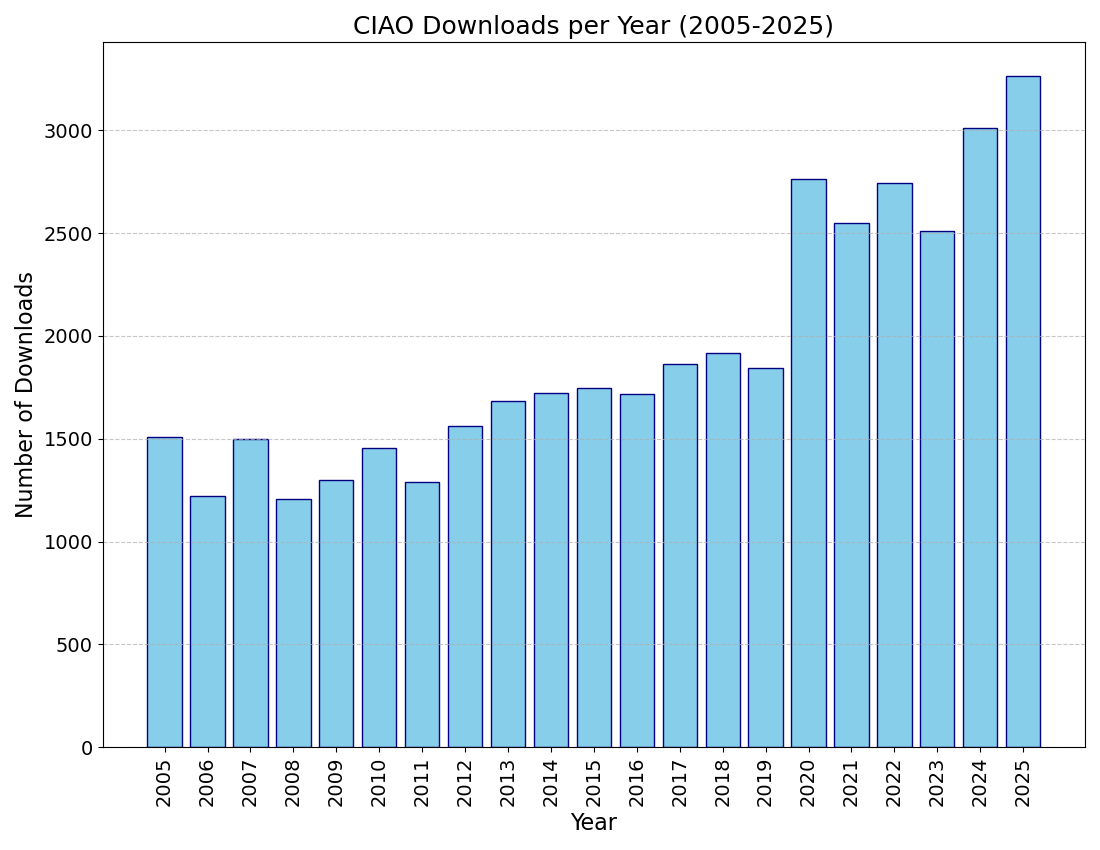}
\caption{Annual number of CIAO downloads until the end of 2025, illustrating sustained usage over time, with thousands of downloads per year. The increase observed since 2020 is likely related to the COVID-19 pandemic, when more researchers started working remotely and installing the software on personal machines. These counts represent downloads, not unique users, so they cannot distinguish new users from existing users upgrading from previous releases.}
\label{fig:downloadsperyear}
\end{figure}

\begin{figure}[ht!]
\includegraphics[width=0.47\textwidth]{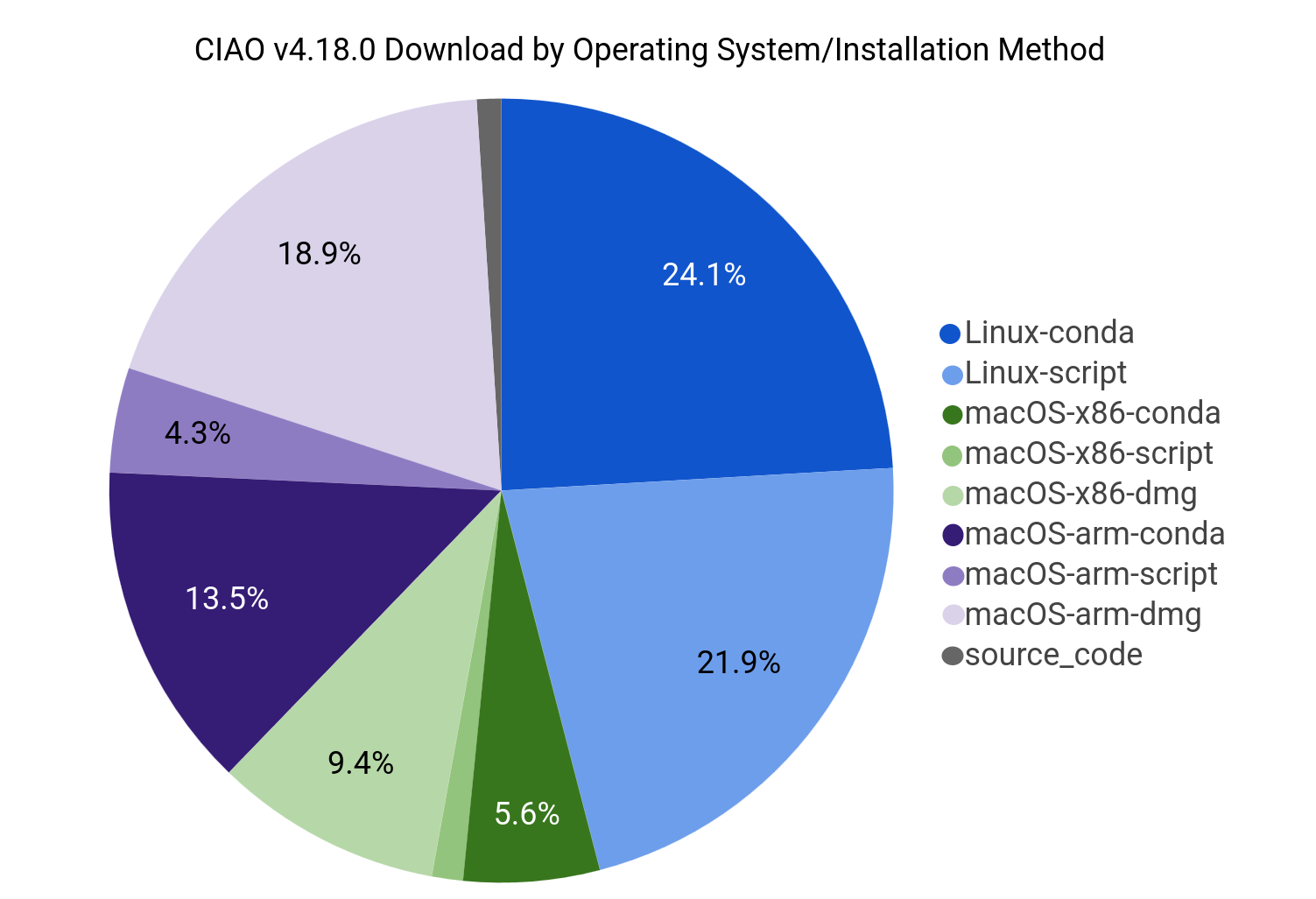}
\caption{Distribution of CIAO v4.18.0 downloads by operating system and installation method, showing that macOS accounts for a slightly larger share of total downloads ($\sim$52\%) than Linux ($\sim$46\%). macOS users are distributed across both x86 and ARM architectures and multiple installation options, while Linux usage is concentrated in \texttt{conda} (24.1\%) and script-based (21.9\%) installations. Source code downloads account for only a small fraction of the total ($<1\%$).}
\label{fig:downloadsbyOS}
\end{figure}

CIAO has evolved into a cross-platform system supporting Linux and macOS. Recent versions utilize the \texttt{conda}\footnote{\url{https://docs.conda.io/projects/conda/en/4.6.0/index.html}} open source package management system, ensuring compatibility across a range of computing environments and hardware. In parallel, development has expanded functionality, introduced many high-level scripts to automate common analysis tasks, and continued development of \sherpa\ to better support multi-dimensional data, forward-fitting techniques, and user-defined models \citep{fruscione2010,siemiginowska2024}. The move to an open development model for \sherpa\ further expanded its applicability and encouraged community contributions. Ongoing development has also emphasized ease of use, support for other X-ray missions, improved tools for PSF characterization---for example subpixel event repositioning methods such as Energy-Dependent Subpixel Event Repositioning (EDSER; \citealt{Li2004})---and statistical tools (e.g.\ \texttt{aprates}, \texttt{lim\_sens} (\citealt{evans2010}, sec~3.8) \citep{fruscione2011_xru,fruscione2024}. Today, CIAO represents nearly three decades of development and remains an important tool in X-ray astronomy. Its continued use is evident in its download statistics, with thousands of downloads per year across all supported platforms (Figures~\ref{fig:downloadsperyear}--\ref{fig:downloadsbyOS}). A bibliography search indicates an average of a few hundred papers per year since 1999 that reference the CIAO system, although this is probably a lower limit given that software usage is often under-cited \citep{bouquin2020}.

\section{Core Features and Technical Capabilities}

CIAO includes tools for users ranging from beginners to advanced. Its toolset has evolved over time to reflect changes in computational methods and user preferences \citep{fruscione2015}. This section summarizes its core capabilities, including modeling, data visualization, and analysis processes.

\subsection{Tools: the Core of CIAO}

At the core of CIAO is a suite of analysis tools that support the full cycle of X-ray data analysis, from raw event processing to high-level scientific interpretation. These tools are designed to be flexible and work together in a consistent analysis system. CIAO includes over 130 core tools, complemented by a large number of high-level scripts.

\subsubsection{The Data Model and Virtual File System}

One of the defining features of CIAO is the CXC Data Model (DM) library, an input/output interface that provides an abstraction layer for handling data in multiple formats. The DM provides a uniform interface to different data types (e.g., event lists, images, and spectra), allowing tools to operate consistently across data products \citep{fruscione2006}. The primary data product in X-ray astronomy is the event file, a FITS\footnote{\url{https://heasarc.gsfc.nasa.gov/docs/heasarc/fits.html}} binary table in which each row corresponds to a detected event or background particle, and records its position, energy, arrival time, and associated metadata. CIAO tools operate directly on these event lists, enabling flexible filtering and binning operations to derive images, spectra, and light curves from a common data representation.

The DM also allows filtering and binning operations to be applied directly within a filename specification, avoiding the need to create intermediate files. This concept of a ``virtual file'' enables on-the-fly data selection without modifying the input dataset \citep{cresitello2007,he2019}. Users can define spatial regions, energy ranges, time intervals, or combinations of these directly in the input string, and CIAO tools operate on the resulting filtered data transparently. This approach extends naturally to binning operations, allowing the creation of $n$-dimensional histograms, such as images or data cubes, from event lists.

To the extent possible, CIAO tools are written in terms of abstract variables rather than explicit references to time, spatial, or spectral axes. The user specifies the relevant variable in the virtual file syntax, allowing the same algorithm to be applied to time series, spectral data, images, or combinations thereof, providing a high degree of flexibility. For example, the \texttt{dmcopy} tool, which writes a filtered or binned virtual dataset into a physical file, can be used with the same syntax to bin different variables depending on the analysis goal: \texttt{dmcopy "evt2.fits[bin sky=::4]" image.fits} creates an image by binning the spatial coordinates, \texttt{dmcopy "evt2.fits[bin time=::100]" lightcurve.fits} creates a lightcurve by binning on time, and \texttt{dmcopy "evt2.fits[bin energy=500:7000:100]" spectrum.fits} creates a spectrum by binning on energy. Variables can also be combined, as in \texttt{dmcopy "evt2.fits[bin time=::100,energy=500:7000:100]" combine.fits}, which bins the data by both time and energy. Another commonly used tool is \texttt{dmlist}, which provides access to data contents, metadata, and file structure. Additional tools such as \texttt{dmextract}, \texttt{dmmerge}, \texttt{dmsort}, and \texttt{dmpaste} provide general-purpose data manipulation.

The DM records filtering operations in the file metadata (the ``data subspace''), ensuring that downstream tools correctly compute quantities such as extraction regions, time intervals, and energy ranges \citep{cresitello2007}. This preserves the analysis history so that processing steps can be traced. CIAO further records parameter values and processing steps in file headers (\texttt{HISTORY} records), allowing analysis to be reproduced.

\subsubsection{Instrument-Specific and Calibration Tools}

Chandra instruments include the Advanced CCD Imaging Spectrometer (ACIS), High Resolution Camera (HRC), and the High Energy and Low Energy Transmission Gratings (HETG and LETG). CIAO instrument-specific tools are tailored to the characteristics of each instrument. For example \texttt{acis\_process\_events} and \texttt{hrc\_process\_events} (to reprocess ACIS or HRC event data by applying instrument-specific calibrations and updating event properties), \texttt{tg\_resolve\_events} (to assign grating events to spectral orders), and \texttt{asphist} (to produce a histogram of the telescope pointing as a function of time) among many. These tools handle instrument-specific corrections, coordinate transformations, and calibration steps required to convert raw data into scientifically usable products.

\begin{figure}[ht!]
\leavevmode\centering
\includegraphics[width=0.48\textwidth]{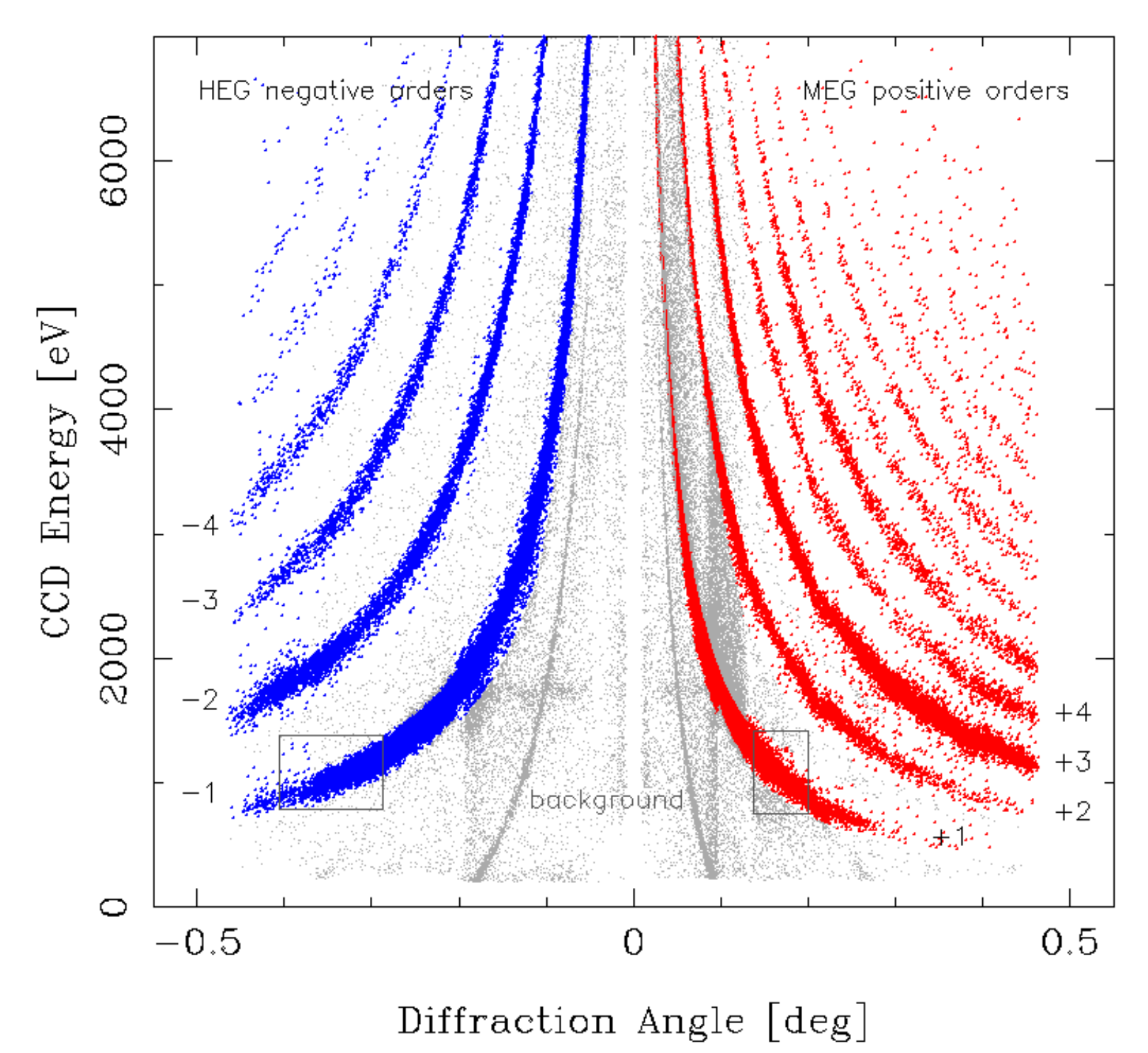}\vspace{2mm}
\includegraphics[width=0.45\textwidth]{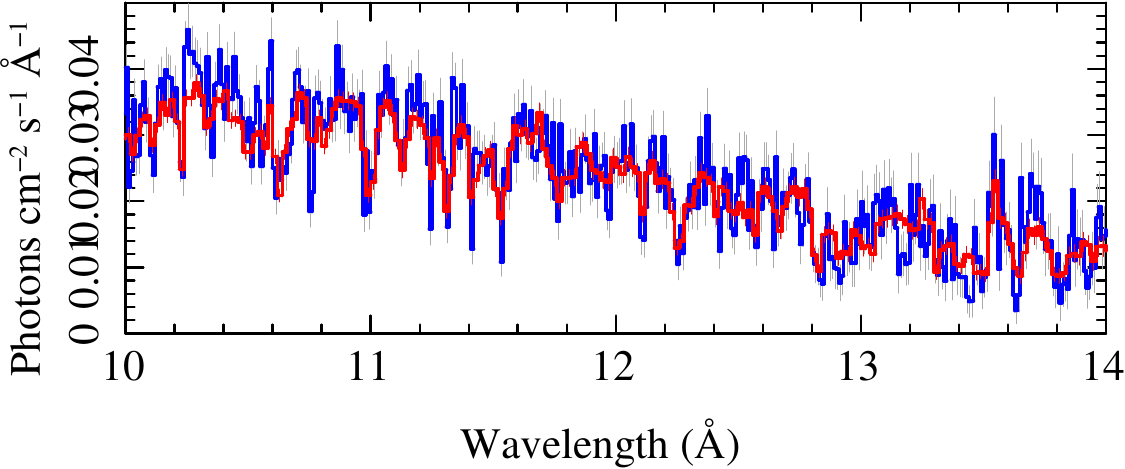}
\caption{Event distribution at an intermediate stage of coordinate transformations in an HETG spectrum. Top: each point represents a detected photon, plotted in ``order-sorting'' coordinates. The $x$-axis shows the diffraction angle (set by where the photon lands on the detector), while the $y$-axis shows the CCD signal converted to an approximate energy. The gratings spread the incoming light into multiple diffraction orders on both sides of the source; here we show HEG negative orders (left, blue) and MEG positive orders (right, red), with orders $1$--$4$ labeled. Photons with different wavelengths can fall at the same angle (same $m\lambda$), but they separate in energy, producing the curved ``banana''-shaped tracks. These tracks allow us to determine which diffraction order each photon belongs to by combining position and energy information. Gray points are background or unresolved events. Bottom: once each photon has been assigned an order and wavelength (e.g., using \texttt{tg\_resolve\_events}), the data can be binned in wavelength and corrected for instrumental effects to produce a spectrum. The higher-resolution curve, with sharper features and shown in blue, is the High Energy Grating (HEG) signal; the smoother red curve is from the Medium Energy Grating (MEG). Data are from Cygnus X\-1 (ObsID 9847).}
\label{fig:hetgorders}
\end{figure}

As an example, Figure~\ref{fig:hetgorders} shows event distributions for an HETG observation at an intermediate stage of processing. The gratings diffract the beam into positive and negative orders; because the High Energy and Medium Energy Gratings (HEG and MEG) have different dispersion directions, their events can be spatially separated (HEG negative orders on the left, MEG positive on the right). The $x$-axis gives the diffraction angle, derived from each event's position, which maps to $m\lambda$, while the $y$-axis shows the CCD energy channel (derived from the PHA). The ``banana''-shaped tracks correspond to different diffraction orders that overlap spatially but separate in energy. At this stage \texttt{tg\_resolve\_events} uses each event's position in conjunction with the diffraction equation to determine $m\lambda$, and uses the CCD energy with the loci allowed by the CCD response function to assign a diffraction order (and hence determine $\lambda$). Now events can be filtered on order, binned in wavelength, and corrected for effective area and exposure to yield the spectra shown in the lower panel.

An important part of this process is the generation and use of calibration products such as ARFs, RMFs, and PSF models. Dedicated tools allow users to create and combine these responses, enabling calibrated spectral and imaging analyses. Integration with the Chandra Calibration Database (CALDB), the repository of all Chandra calibration data, provides access to up-to-date calibration data throughout the analysis \citep{graessle2006}. Calibration files are selected dynamically based on observation metadata such as observation date, detector configuration, and operating mode, ensuring that analyses use the most appropriate and up-to-date calibration products.

\subsubsection{General Analysis Capabilities}

\begin{figure}[ht!]
\includegraphics[width=0.47\textwidth]{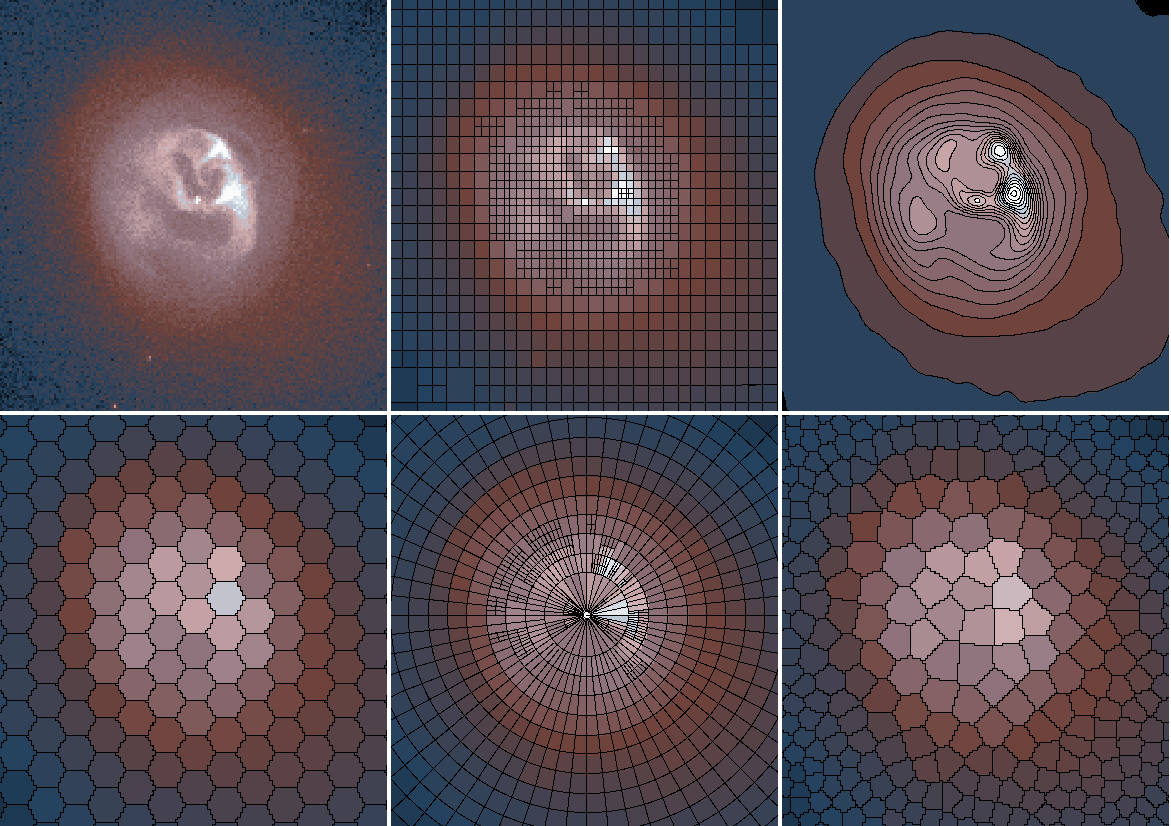}
\caption{Demonstration of several adaptive and alternative binning approaches available in CIAO. All images show a 620 ks observation of Abell 2052 in the 0.5--7.0 keV band. The top left panel shows the original image with standard pixel binning (each pixel is $\sim 1^{\prime\prime}$ on a side). The left column compares uniform tilings, including square pixels (top) and hexagonal bins (bottom, generated with \texttt{hexgrid}, side length $\sim 4^{\prime\prime}$). In the center column, the images are adaptively binned to have at least 900 counts per bin: rectilinearly (top) using \texttt{dmnautilus} and in polar coordinates (bottom) using \texttt{dmradar}. The right column shows more morphologically complex binning: a contour map (top) generated with \texttt{dmcontour} and \texttt{mkregmap} and a centroid map (bottom) produced by \texttt{centroid\_map}, which is based on Voronoi tessellation of the local maxima in the image. Different binning methods can be necessary when exploring properties e.g. temperature or hardness ratio maps of large extended objects, such as galaxy clusters.}
\label{fig:binning}
\end{figure}

In addition to instrument-specific tools, CIAO provides general-purpose tools for a wide range of astrophysical analyses. These include source detection (e.g., \texttt{wavdetect} (based on a wavelet transform algorithm); \citealt{freeman2002} and \texttt{vtpdetect} (based on the Voronoi Tessellation and Percolation); \citealt{ebeling1993}), image processing (smoothing, e.g.\ \texttt{csmooth}; \citealt{ebeling2006}, adaptive binning, reprojection, and regridding), region generation (e.g., \texttt{dmcontour}, \texttt{dmimglasso}, \texttt{dmimghull}), timing analysis (e.g., \texttt{glvary}; \citealt{gregory1992}, \texttt{pfold}, \texttt{dmgti}), statistical analysis (e.g., \texttt{dmstat}, \texttt{aprates}; \citealt{primini2014}), and metadata and parameter management (e.g., \texttt{dmhedit}, \texttt{dmhistory}). Because these tools share a common design, they can easily be combined into robust analysis workflows. Examples of advanced imaging functionality include adaptive and morphology-driven binning techniques, as illustrated in Figure~\ref{fig:binning}.

\subsubsection{Design Philosophy and Usability}

The CIAO toolset is built around a command-line interface with parameter files, supporting both interactive and scripted use. This allows users to run individual tools or combine them into automated pipelines, while maintaining a consistent analysis environment \citep{burke2006}.

An important aspect of this design is that the same tools are used both in the Chandra data processing pipeline and by end users. As a result, data products in the archive and those generated by users are processed in a consistent way, making it easier to compare results and understand the processing steps applied \citep{mcdowell2006}. Since calibration products and algorithms are updated independently of the archive reprocessing, users can also apply the most recent calibration files and methods to their data before updated versions are available in the archive.

Features such as parameter prompting, command history, and integrated help (``ahelp'') make the system easier to use, while its design, where each tool is developed to perform a single, well-defined task, provides flexible, universal building blocks that can be combined in different ways to support advanced analyses.

Overall, the CIAO tools provide a consistent and extensible system for data analysis, combining flexible data handling, calibration support, and a wide range of scientific capabilities. By adhering to community standards for high-energy astrophysics data formats (e.g., OGIP for event lists, spectra, and response files), CIAO maintains interoperability with software and data from other missions.

\section{High-Level Scripts: Enhancing Productivity}

CIAO includes an extensive set of ``contributed scripts''\footnote{\url{https://cxc.harvard.edu/ciao/download/scripts/index.html}} that extend the functionality of the core tools and simplify common analysis tasks. The term ``contributed'' is historical, reflecting their origin as user-supplied scripts in multiple languages; today they are implemented in Python and maintained by the CXC. These scripts automate repetitive tasks and provide higher-level interfaces to complex operations, helping users carry out analysis more efficiently across a wide range of applications \citep{fruscione2023}. They are accompanied by detailed documentation and analysis threads, allowing new users to understand the data analysis procedure, while their modular design allows advanced users to customize or extend them as needed.

\subsection{Coverage of the Contributed Scripts}

\begin{figure}[ht!]
\includegraphics[width=0.47\textwidth]{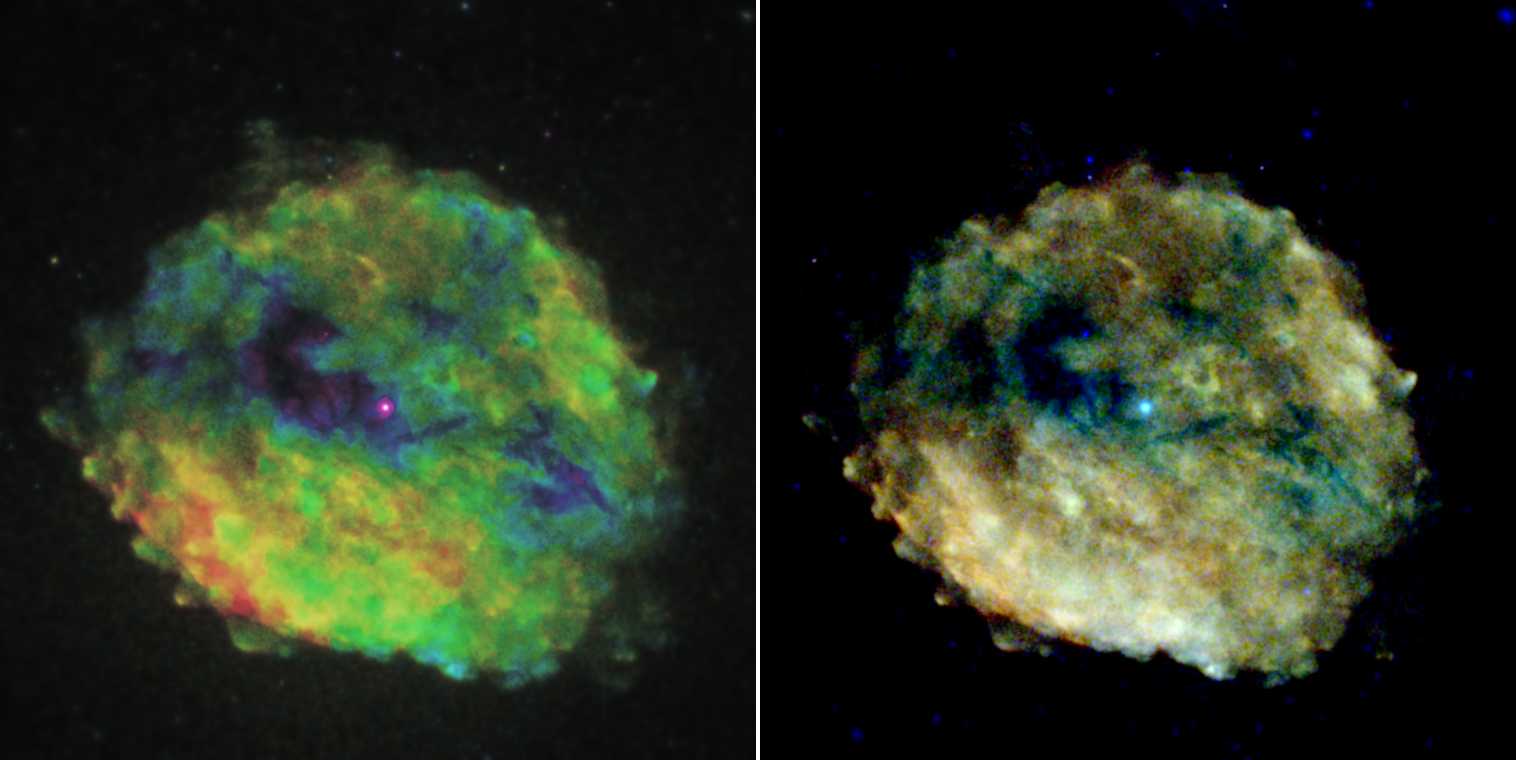}
\caption{X-ray image of the supernova remnant RCW 103 using two methods of color representation: true color (left) and tri-color (right). The true color image was created using the \texttt{energy\_hue\_map} script, which combines the adaptively smoothed counts image with the median energy at each location to create a true color image. Here, the color hues span a continuous range capped by red (median energies below 1.0 keV) and violet (median energies above 1.4 keV). The tri-color image was created using standard Chandra Source Catalog energy bands: soft (0.5--1.2 keV), medium (1.2--2.0 keV), and hard (2.0--7.0 keV). Traditional Red-Green-Blue (RGB) tri-color images start by creating 3 images in separate energy bands, which are independently scaled to act as the individual color channels of the output image. The combination of primary colors (RGB) leads to the visual interpretation of secondary colors (e.g., yellow, cyan etc.). True color images, in contrast, are created from a continuous energy range producing images in the Hue-Saturation-Value (HSV) color system.}
\label{fig:energyhuemap}
\end{figure}

The contributed scripts cover all major stages of X-ray data analysis. These include data retrieval from the Chandra archive (e.g., \texttt{download\_chandra\_obsid}), data preparation and reprocessing---for example to apply updated calibration files or use different analysis parameters than in standard processing---(e.g., \texttt{chandra\_repro}, \texttt{splitobs}), imaging analysis (e.g., \texttt{fluximage}), source characterization (e.g., \texttt{src\_psffrac}), grating analysis for HETG and LETG observations (e.g., \texttt{mktgresp}, \texttt{combine\_grating\_spectra}), PSF modeling and visualization (e.g., \texttt{simulate\_psf}), and general utilities for data manipulation and environment verification (e.g., \texttt{check\_ciao\_caldb}). Python modules provide access to CIAO tools, allowing them to be integrated into custom workflows and Python-based analysis environments. An example of an advanced visualization option is shown in Figure~\ref{fig:energyhuemap}, where the \texttt{energy\_hue\_map} script is used to create a true-color representation of X-ray data. Together, these scripts support logical workflows from data acquisition and preparation to scientific interpretation.

\subsection{Integration with CIAO}

The contributed scripts are tightly integrated with the core CIAO environment, including \sherpa\ for modeling and SAOImageDS9 for visualization, or ChaRT and MARX (see section \ref{sec:sim}) for simulations. Although implemented as higher-level wrappers, they operate within the standard CIAO analysis framework---utilizing the same parameter interface---allowing users to move smoothly between data preparation and scientific analysis \citep{fruscione2023}.

\subsection{Evolution and Impact}

\begin{figure}[ht!]
\includegraphics[width=0.47\textwidth]{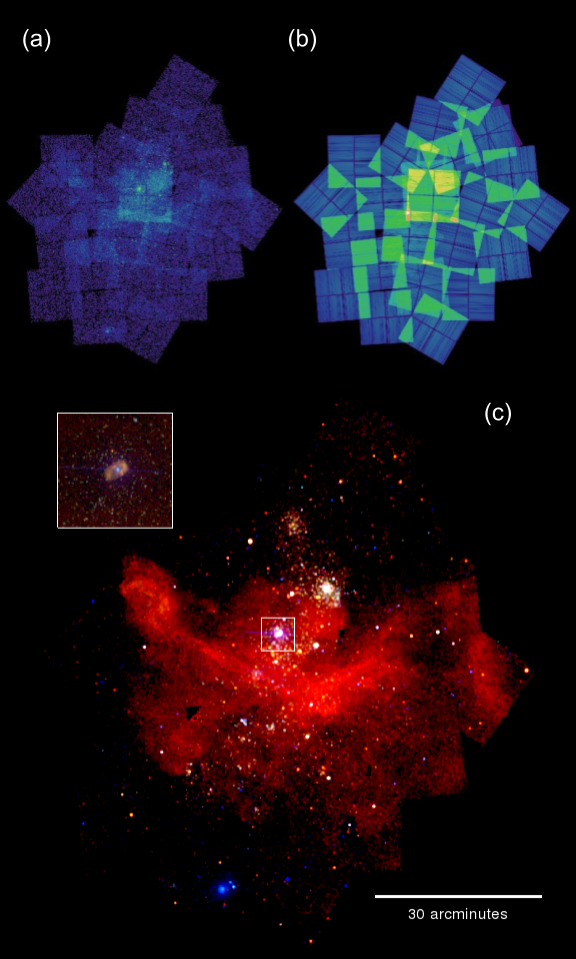}
\caption{A mosaic of 41 Chandra ACIS-I observations centered near $\eta$ Carinae, processed using the \texttt{merge\_obs} script. (a) Integrated 0.5--7.0 keV counts image after all observations have been reprojected to a common tangent point. (b) The corresponding broadband monochromatic exposure map (in $\text{cm}^2\,\text{s}\,\text{count}\,\text{photon}^{-1}$), used to account for spatial variations in instrument sensitivity and vignetting. (c) A fluxed true-color image created by dividing the reprojected data by the exposure maps in the soft (0.5--1.2 keV), medium (1.2--2.0 keV), and hard (2.0--7.0 keV) bands, followed by smoothing. The inset provides a high-resolution view of the $\eta$ Carinae nebula.}
\label{fig:etacarinae}
\end{figure}

The contributed scripts package has evolved in response to user feedback and scientific needs, with several additions driven by recurring analysis challenges and community-developed solutions later incorporated into CIAO. For example, the ``blank-sky background'' workflow was initially user-developed, then distributed as a thread, and later implemented as the \texttt{blanksky} script. The original \texttt{merge\_all} script similarly evolved into the more robust \texttt{merge\_obs}, incorporating user-driven improvements to reproject and combine multiple observations, and to produce merged event files and exposure-corrected images. An example of this script is demonstrated in Figure~\ref{fig:etacarinae}, which shows a mosaic of Chandra observations. The script automates the complex process of reprojecting different observations to a common tangent point, creating exposure maps, and generating fluxed, multi-band images that account for spatial variations in sensitivity.

\begin{figure}[ht!]
\includegraphics[width=0.47\textwidth]{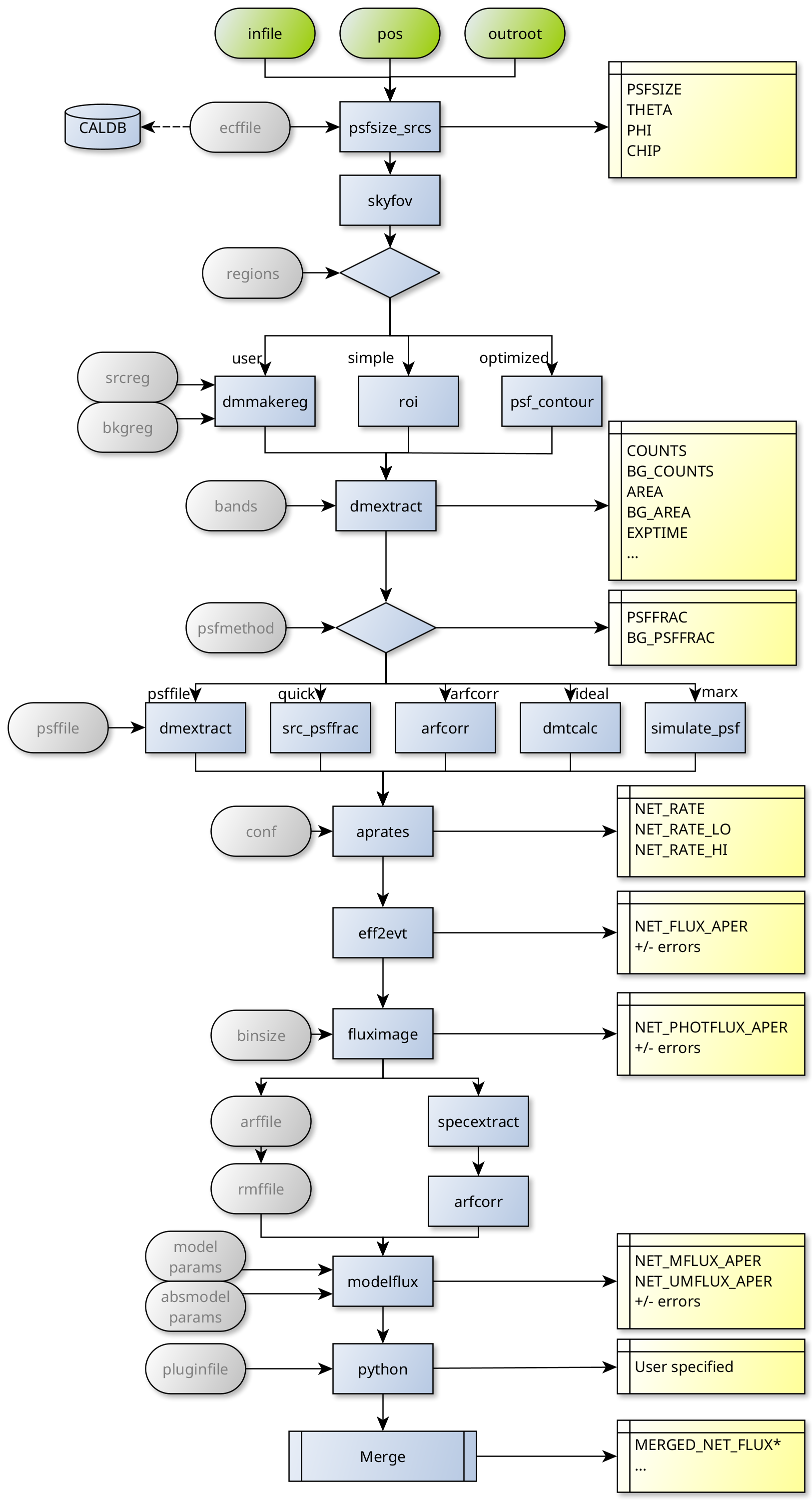}
\caption{Schematic diagram of the \texttt{srcflux} script illustrating the complexity of the analysis pipeline. The procedure links multiple CIAO tools and intermediate data products, starting from an input event file, source position, and output root, and proceeding through region definition, counts extraction in multiple energy bands, and background estimation. Several alternative branches are used to compute PSF corrections, depending on the selected method, before combining these quantities to derive net count rates with \texttt{aprates}. Subsequent steps convert these rates into energy and photon fluxes, with optional spectral extraction and model-based flux calculations. The diagram highlights the numerous dependencies, conditional paths, and intermediate products involved, illustrating how \texttt{srcflux} encapsulates a complex, multi-step analysis into a single high-level script.}
\label{fig:srcflux}
\end{figure}

The high-level scripts frequently convert multi-step analysis threads into streamlined processes. For example, the \texttt{srcflux} script (\citealt{glotfelty2014} and Figure \ref{fig:srcflux}), inspired in part by \texttt{acis\_extract} \citep{broos2010}, combines a large number of core tools and scripts into a single interface for source flux estimation. Likewise, \texttt{fine\_astro} integrates functionality from \texttt{merge\_obs}, \texttt{wavdetect}, and \texttt{reproject\_aspect} to automate astrometric alignment.

\section{sherpa: Modeling and Fitting}

\begin{figure}[ht!]
\includegraphics[width=0.47\textwidth]{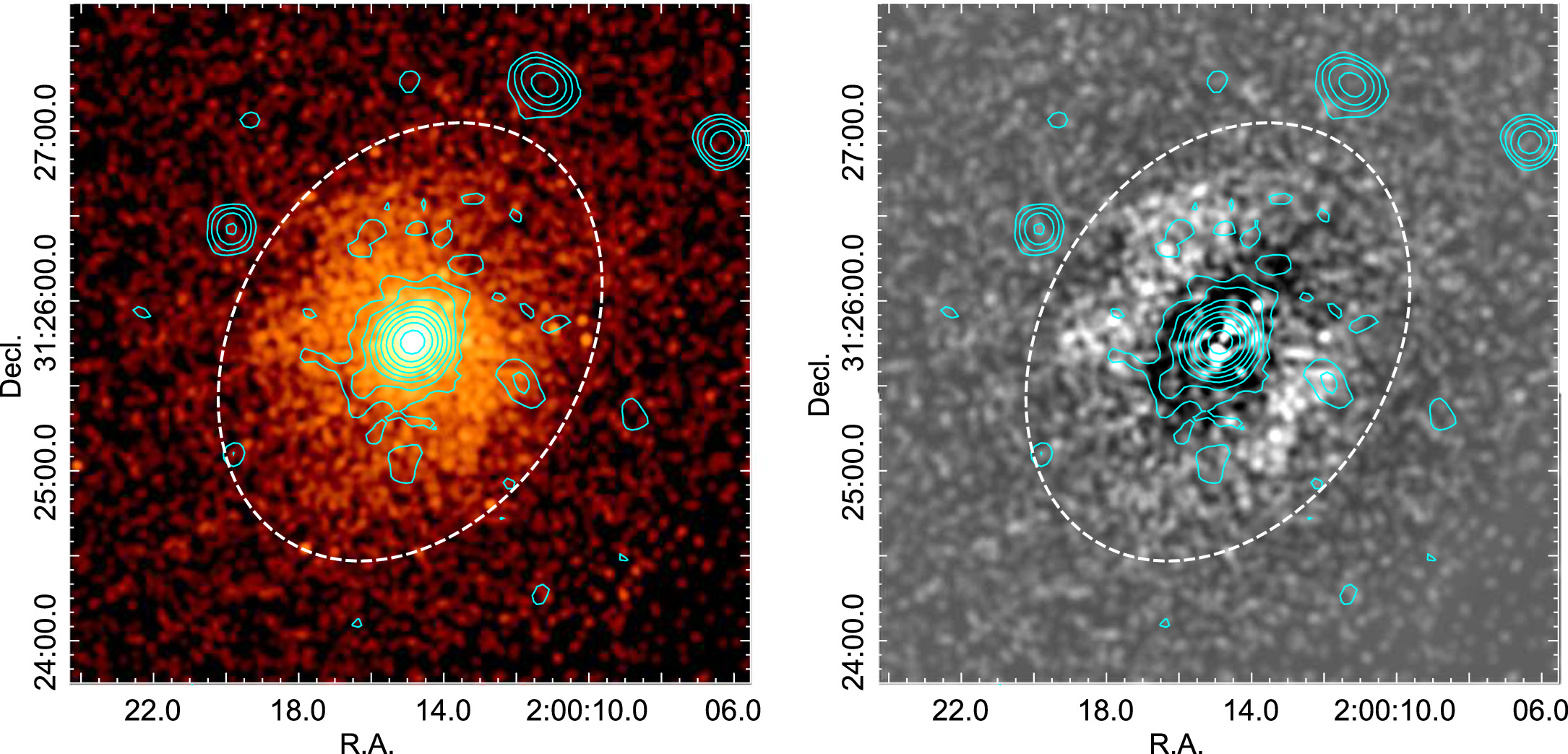}
\caption{Example of two-dimensional modeling of the X-ray surface brightness of NGC 777 using \sherpa\ \citep{osullivan2024}. Left: Chandra 0.5--2 keV exposure-corrected image, smoothed with a $2.5^{\prime\prime}$ radius Gaussian. Cyan contours indicate the 400 MHz radio emission. The dashed ellipse shows the approximate $D_{25}$ contour of the galaxy. Right: Residual map after the removal of point sources and subtraction of the best-fitting two-dimensional surface brightness model, consisting of an elliptical $\beta$ model describing the large-scale intragroup medium emission and a central Gaussian component representing the compact core. The residuals highlight excess emission to the northeast and southwest of the core and a deficit along the northwest--southeast axis, revealing deviations from a smooth, azimuthally symmetric surface brightness distribution.}
\label{fig:sherpa2D}
\end{figure}

\sherpa\ is an open-source Python modeling and fitting application\footnote{\url{https://github.com/sherpa/sherpa}} \citep{siemiginowska2024}. It incorporates forward-fitting techniques to fit complex models using robust statistical methods and optimization routines \citep{freeman2001, refsdal2009}. Its library of parametric and non-parametric models supports the analysis of spectra (including XSPEC spectral models), images (see Figure~\ref{fig:sherpa2D}), and time-series data, with applications across a range of problems in multiwavelength astronomy and even non-astronomical datasets.

\sherpa\ supports high-complexity modeling, including the combination of parametric models into sophisticated expressions and the incorporation of ancillary data such as PSFs, background maps, and exposure maps in 2D image models to ensure accurate scientific inference. \sherpa\ incorporates Bayesian inference, leveraging Markov Chain Monte Carlo (MCMC) methods to sample parameters and compute posterior distributions \citep{vandyk2001, protassov2002}. These capabilities are especially important in high-energy astrophysics, where the data are inherently Poisson-distributed and require appropriate statistical treatment. They are further supported by developments from the CHASC International AstroStatistics Center\footnote{\url{https://hea-www.harvard.edu/AstroStat/}}, an interdisciplinary collaboration between astrophysicists and statisticians focused on statistical methods for astrophysical data analysis.

\begin{figure}[ht!]
\includegraphics[width=0.47\textwidth]{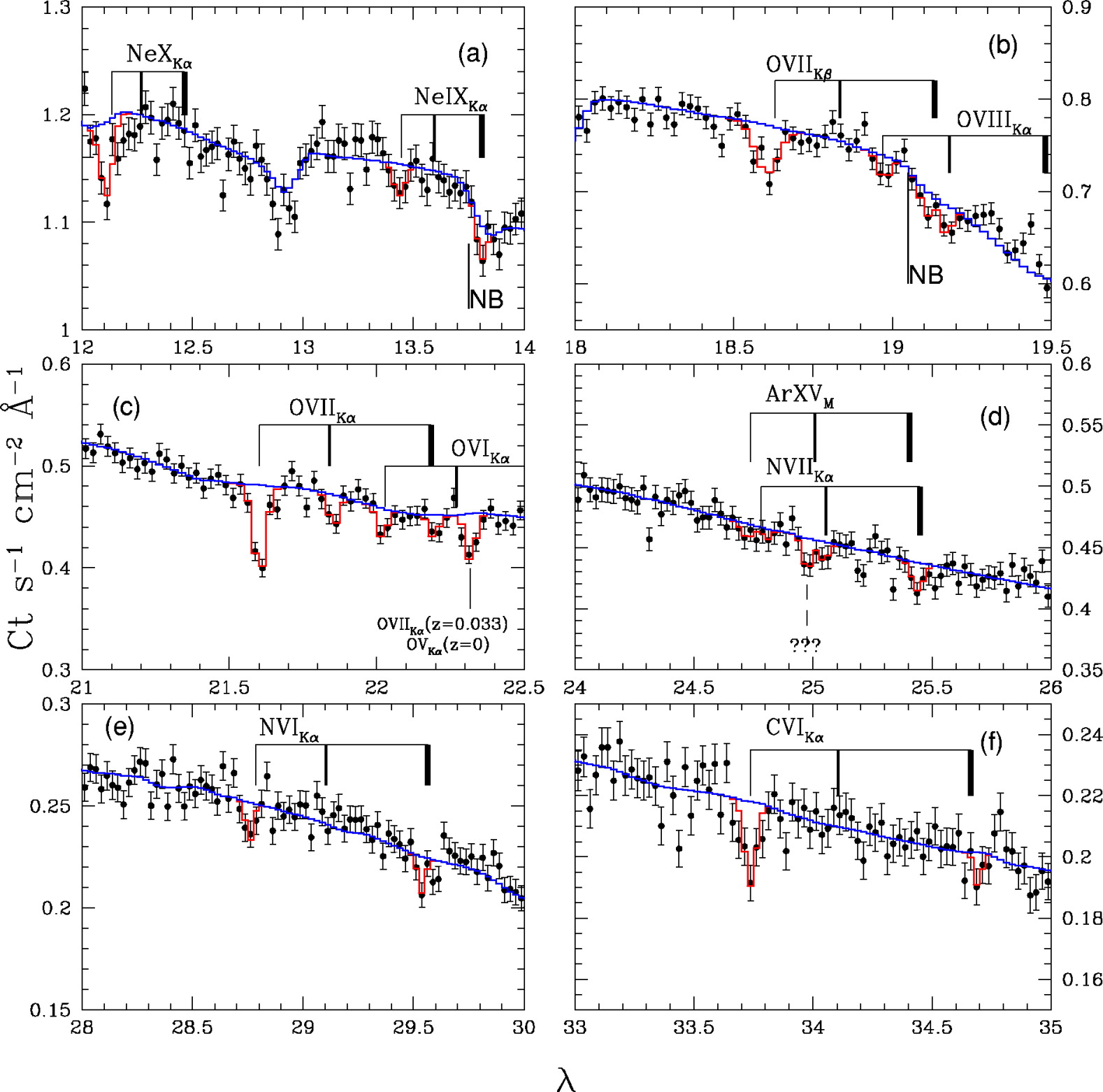}
\caption{High-resolution LETG spectra of the blazar Mrk~421 showing absorption features from the warm--hot intergalactic medium (WHIM) \citep{nicastro2005}. The panels display selected wavelength regions with lines from highly ionized species (e.g., Ne~X, O~VII, O~VIII). Black points represent the data; the model is obtained via forward fitting with instrumental response folding, with the blue curve indicating the continuum and red histograms the absorption components. Fitting multiple weak and partially blended lines in low-count data requires an appropriate Poisson statistical framework. This example highlights the complexity of high-resolution X-ray spectroscopy and demonstrates \sherpa's ability to model multi-component absorption systems relevant to studies of the X-ray forest and intergalactic baryons.}
\label{fig:letg}
\end{figure}

An example of high-resolution spectral modeling with \sherpa\ is shown in Figure~\ref{fig:letg}, based on the analysis of LETG grating spectra of Mrk~421 \citep{nicastro2005}, where multiple absorption features are fit simultaneously. It highlights \sherpa's ability to model low-count data with complex, multi-component models.

\sherpa's open development transition in 2014 marked a significant milestone, enabling greater community involvement in its development and ensuring compatibility with current computational environments. Hosted on GitHub, \sherpa\ has released over 20 public versions and is now available as part of CIAO or as a standalone Python package. Its design supports both interactive use, such as within Jupyter Notebooks, and integration into larger analysis pipelines (e.g., in the processing of the Chandra Source Catalog).

\sherpa's adoption spans a variety of research domains. It has been used for modeling X-ray spectra of active galactic nuclei (AGN), fitting optical spectra from the Sloan Digital Sky Survey (SDSS) and Gemini \citep{leighly2025} and analyzing multiwavelength data from observatories like Hubble, Spitzer, and NuSTAR \citep{nigro2022, ilic2023}. It has also been a critical tool for large-scale projects, such as the Chandra Source Catalog, where \sherpa\ was employed to fit over one million images and source detections \citep{evans2010, evans2024}. The combination of robust methodology, flexibility, and community-driven development ensures that \sherpa\ remains a key tool in advancing astrophysical research.

\section{Visualization: Imaging and Plotting}

\subsection{SAOImageDS9}

SAOImageDS9 (DS9; \citealt{joye1999,joye2003,glotfelty2022,fruscione2026}) serves as the primary visualization interface within the CIAO system and is particularly well suited for the analysis of X-ray event data. DS9 can directly display event files and supports several coordinate systems, allowing users to explore photon distributions, apply spatial and energy filtering, and interactively define analysis regions.

\begin{figure}[ht!]
\includegraphics[width=0.47\textwidth]{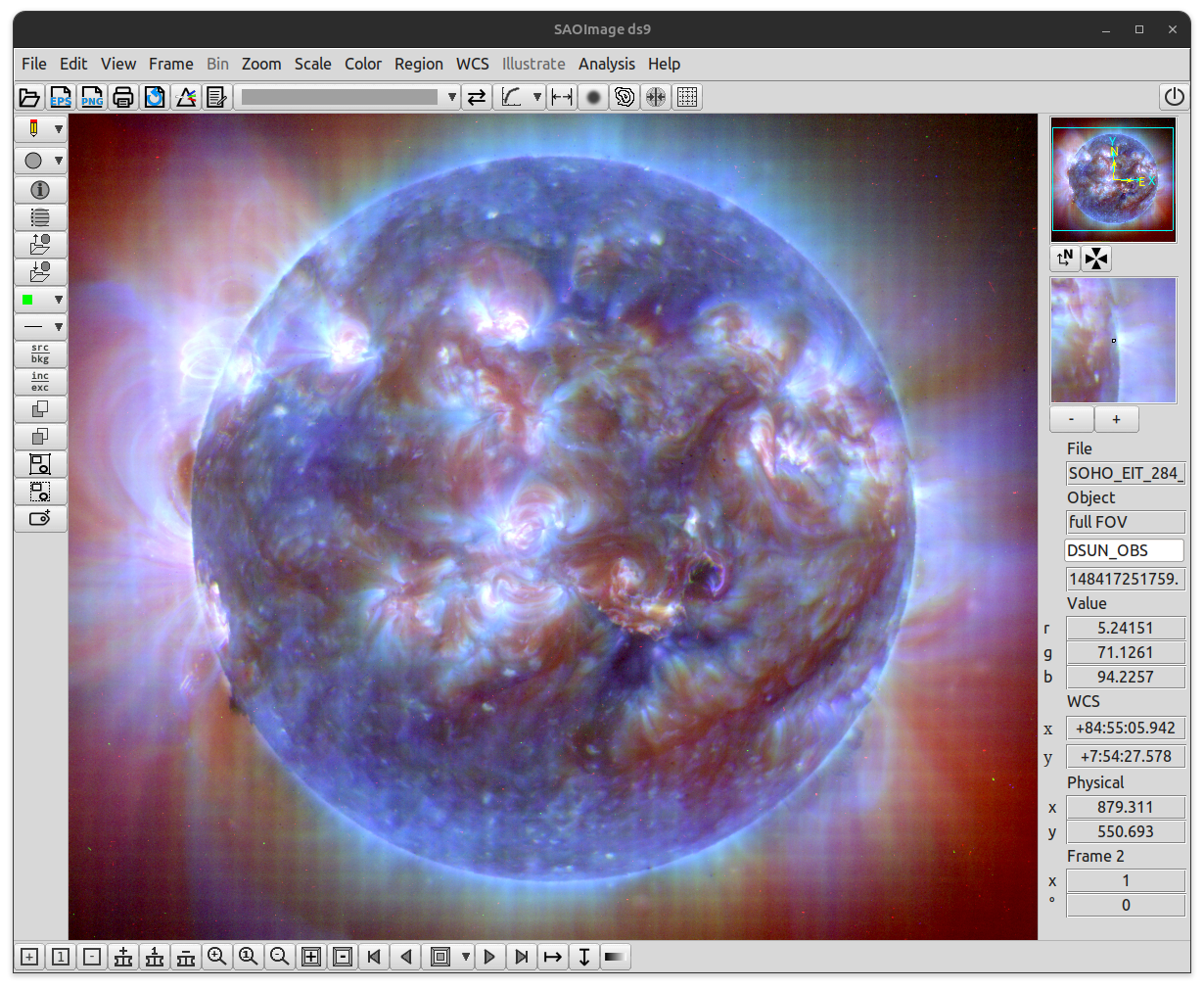}
\caption{Example of SAOImageDS9 in the ``advanced'' view displaying solar data from the Solar and Heliospheric Observatory (\textit{SOHO}). The image is an RGB composite of extreme ultraviolet observations (in three wavelengths: red = 284~\AA, green = 195~\AA, blue = 171~\AA), highlighting structures in the hot solar atmosphere. DS9 supports solar coordinate systems through integration with the AST world coordinate system library \citep{berry2012}, enabling accurate interpretation of helioprojective coordinates.}
\label{fig:ds9}
\end{figure}

In addition to visualization, DS9 allows direct access to the FITS file structure, including headers and table data, and supports interactive inspection of the data by incorporating the functionality previously provided by the deprecated CIAO \textit{Prism} graphical interface \citep{glotfelty2022}. This integration provides a unified environment for both visualization and data exploration.

The CIAO DS9 Analysis eXtensions (DAX) further enhance this capability by connecting to selected CIAO tools which can be executed directly within the DS9 interface \citep{glotfelty2011, glotfelty2020}. These extensions provide an easy introduction to CIAO for new users and are particularly helpful when analyzing the multi-wavelength data, including optical or radio wavebands in addition to X-rays. This connection between visualization and analysis allows for convenient transition between interactive exploration and quantitative analysis, making DS9 an essential part of the CIAO system.

Although DS9 is deeply integrated within the CIAO environment and widely used for the analysis of Chandra data, it is not limited to X-ray astronomy. DS9 supports data from a broad range of ground- and space-based observatories across the electromagnetic spectrum, allowing multiwavelength analysis within a single interface. It can display and interact with diverse data products, including images, event lists, and data cubes, allowing observations from different instruments and wavebands to be compared and combined. In particular, multiple frames can display data from different wavebands (e.g., radio and X-ray), which can be aligned using WCS information to share a common coordinate system, enabling direct comparison and measurement of features across datasets. An example is shown in Figure~\ref{fig:ds9}, where DS9 is used to display solar data from the SOHO spacecraft. This flexibility has led to its widespread adoption beyond the X-ray community, establishing DS9 as a general-purpose astronomical visualization and analysis application, with tens of thousands of downloads per year.

\begin{figure}[ht!]
\includegraphics[width=0.47\textwidth]{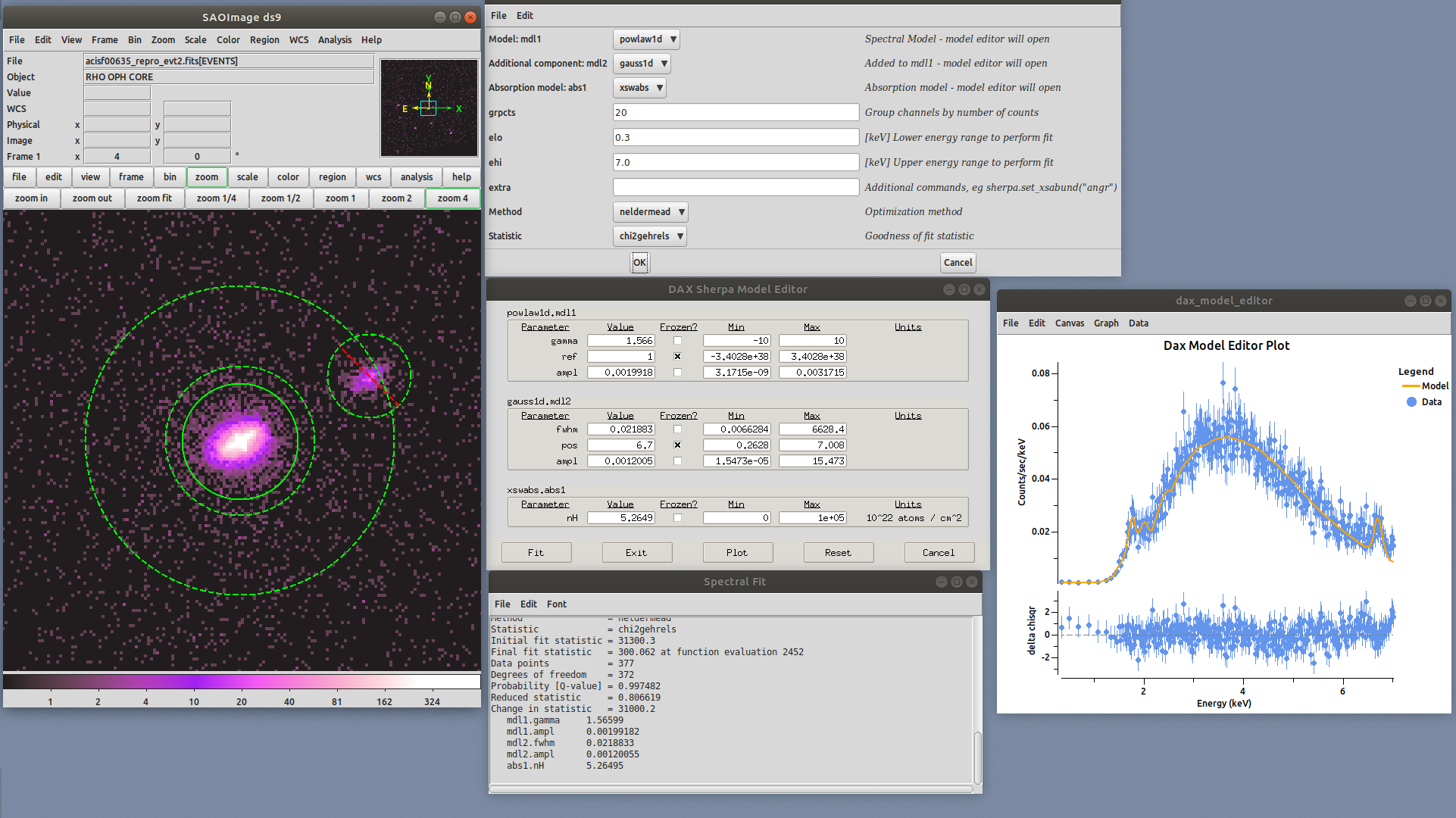}
\caption{An example of the capabilities of DAX showing the \texttt{Spectral Fit} analysis task, which extracts a Chandra spectrum, including response files, and fits the data with Sherpa. This task highlights many DAX features. The SAOImageDS9 window (left) displays source and background regions: a solid green circle for the source, a dashed annulus for the background, and a nearby source excluded from the background (dashed circle with a red line across). The top center panel shows analysis parameters, including model selection, energy range, grouping, and fit statistic. The middle panel shows the DAX Model Editor, a Python-based interface (a Tkinter widget) for adjusting model parameters and limits. The bottom panel shows the text output from Sherpa as the fit is performed. The right panel shows the spectrum (blue points) with the best-fit model (orange line) and residuals below.}
\label{fig:dax}
\end{figure}

\subsection{Plotting: from ChIPS to Matplotlib}

For over a decade, the Chandra Interactive Plotting System (\textit{ChIPS}) was the primary plotting tool within CIAO, providing capabilities for interactive visualization and the creation of publication-quality figures \citep{miller2015}. As user preferences shifted towards Python-based visualization tools, development of \textit{ChIPS} was frozen in later CIAO releases and the tool was subsequently removed, reflecting the broader transition of the scientific community \citep{fruscione2017}.

This transition was accompanied by the adoption of open source \textit{Matplotlib} \citep{hunter2007}, aligning CIAO with the broader scientific Python ecosystem and enabling seamless integration of visualization within data analysis threads. Documentation was provided to facilitate the migration of existing pipelines, ensuring continuity for users. The adoption of \textit{Matplotlib} represents a significant update of CIAO plotting capabilities, providing a familiar and widely supported environment for visualization.

\section{\textit{Chandra} Simulations}\label{sec:sim}

CIAO is complemented by a set of simulation tools and applications that allow detailed modeling of the Chandra instrument response and support the interpretation of observational data. These applications are particularly important for understanding instrumental effects, validating analysis methods, and comparing observations with theoretical models.

The Chandra Ray Tracer\footnote{\url{https://cxc.harvard.edu/ciao/PSFs/chart2/index.html}} (ChaRT; \citealt{carter2003}) is a web interface to SAOTrace\footnote{\url{https://cxc.cfa.harvard.edu/cal/Hrma/SAOTrace.html}}, the raytrace model which simulates the propagation of X-ray photons through the Chandra optics \citep{jerius2004}. ChaRT provides high-fidelity simulations of the telescope PSF, accounting for the complex mirror geometry and energy-dependent behavior of the system. These simulations are commonly used to generate the detailed PSF models for comparison with observed data.

\begin{figure}[ht!]
\includegraphics[width=\linewidth]{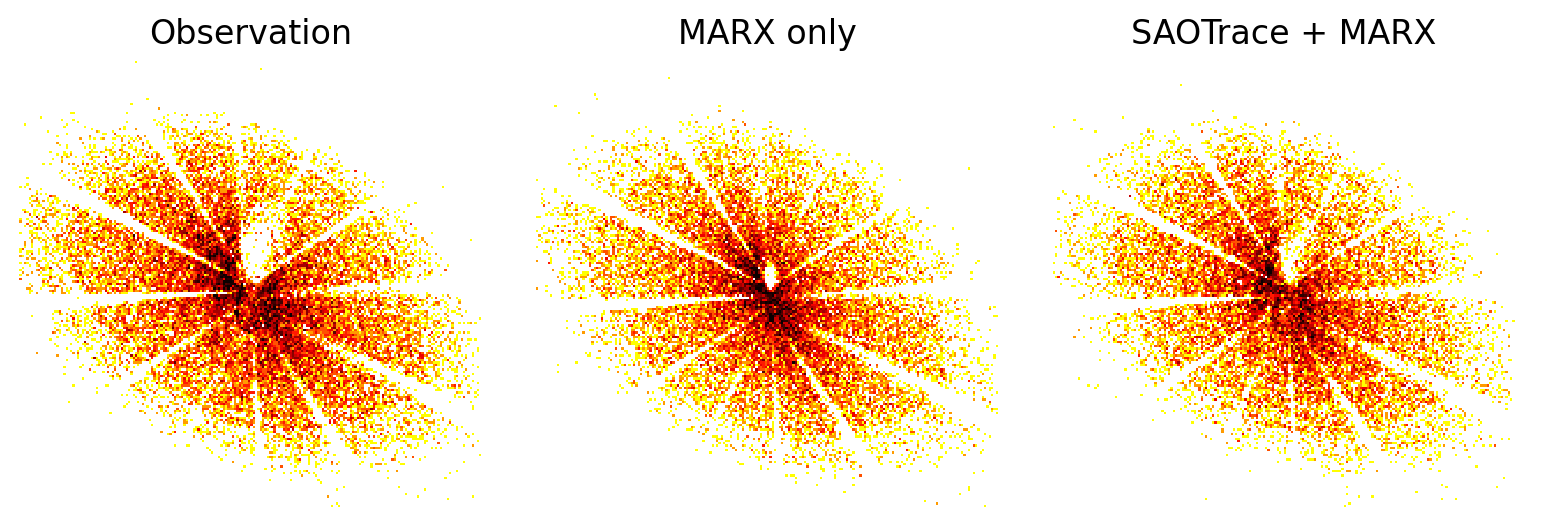}
\caption{Observed PSF of a bright off-axis Chandra calibration source (left; LMC~X-1 observed $\sim25^{\prime}$ off-axis with the ACIS detector), compared with simulations using MARX only (center) and SAOTrace+MARX (right). All three images show remarkably similar large-scale PSF structure, demonstrating the overall accuracy of both mirror models. A closer comparison reveals a subtle asymmetry just above and to the right of the strut-shadow intersection. This feature is slightly less pronounced in the MARX-only simulation, reflecting its simplified mirror model, while SAOTrace reproduces it more faithfully. In practice, such fine differences are rarely significant, as few sources are observed with sufficient depth at such large off-axis angles.}
\label{fig:psfsim}
\end{figure}

The MARX\footnote{\url{https://chandra-marx.github.io/}} (Model of AXAF response to X-rays) software package extends the capabilities of ChaRT by modeling the interaction of photons with the instrument detectors \citep{davis2012}. Compared to ChaRT, MARX employs a simplified but significantly faster ray trace of the Chandra mirrors, assuming idealized optics and incorporating mirror imperfections and scattering through statistical parameterizations. Comparisons show good agreement with SAOTrace for off-axis sources (beyond $\sim5'$), while differences become apparent in the detailed structure of the on-axis PSF, particularly in the wings (see Appendix in \citealt{primini2011}). As illustrated in Figure~\ref{fig:psfsim}, the simulated PSFs reproduce the observed off-axis structure well.

\begin{figure}[ht!]
\includegraphics[width=0.47\textwidth]{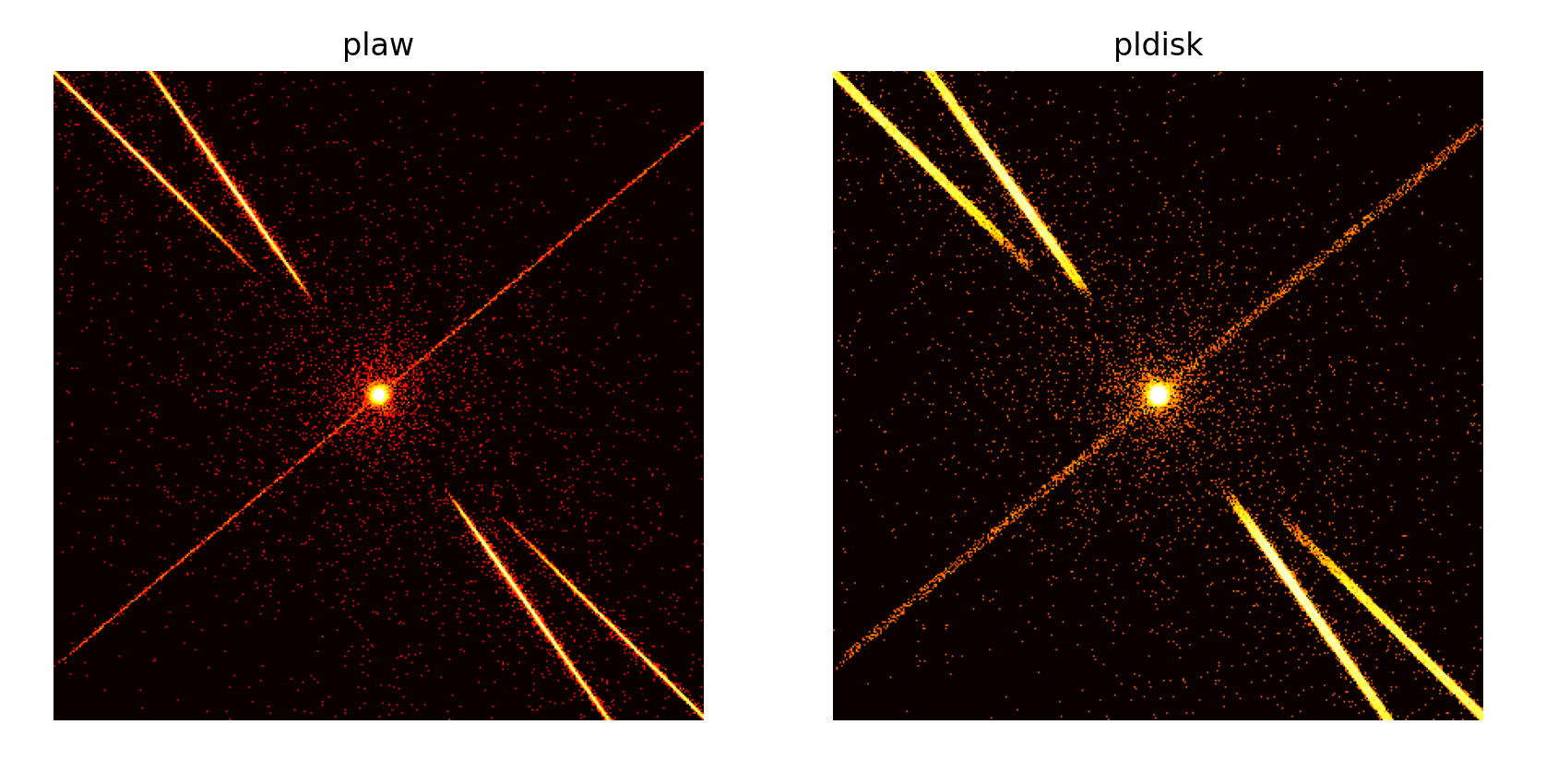}
\caption{MARX simulations of Chandra HETG observations for two source models: a point source with a power-law spectrum (``\texttt{plaw}'', left) and an extended disk-like source (``\texttt{pldisk}'', right). In both cases, the zeroth-order image is accompanied by dispersed grating arms corresponding to multiple diffraction orders. The simulations reproduce the spatial distribution of photons along the dispersion directions, as well as the morphology of the source, demonstrating MARX's capability to model full grating observations, including both imaging and spectroscopic components of the instrument response.}
\label{fig:hetgsim}
\end{figure}

MARX can take as input either a source model or the high-fidelity output of ChaRT, and produces simulated event files that can be analyzed using standard CIAO tools, enabling direct comparison between simulated and observed datasets. An example of grating simulations is shown in Figure~\ref{fig:hetgsim}, demonstrating its ability to model both imaging and dispersed spectral components.

The CIAO script \texttt{simulate\_psf} provides a high-level interface to these simulation tools, simplifying the generation of PSF simulations for existing observations. It uses the observation metadata to configure the simulation and projection, and can operate either directly with MARX or with rayfiles generated by ChaRT. In the latter case, the rays can be projected onto the detector using MARX or the CIAO tool \texttt{psf\_project\_rays}, allowing flexible integration of ray tracing with detector simulations.

\begin{figure}[ht!]
\includegraphics[width=\linewidth]{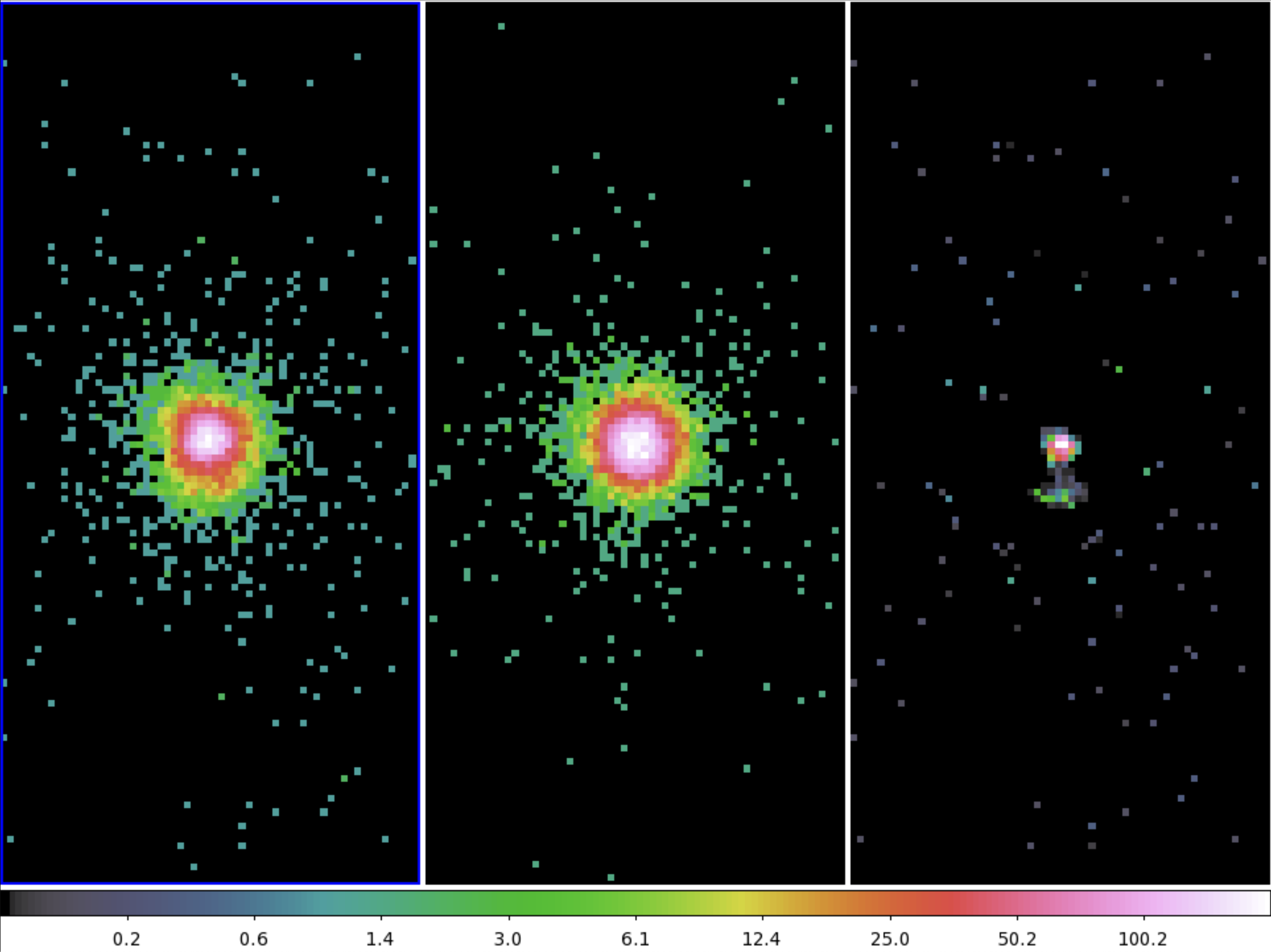}
\caption{On-axis point-source simulation and PSF-based image reconstruction for $\tau$~Canis~Major observed with the ACIS detector (ObsID 4469; ACIS-I, NGC~2362). \textit{Left:} observed image. \textit{Center:} MARX-simulated PSF generated using \texttt{simulate\_psf}. \textit{Right:} image restored via deconvolution with the source PSF using the \texttt{arestore} tool and the Lucy--Richardson method. All three images are shown on the same angular scale. The overall structure of the source is well reproduced, while the deconvolved image reveals a small feature to the south of the source. This feature is not astrophysical but likely reflects a known asymmetry in the Chandra High Resolution Mirror Assembly (HRMA) PSF\footnote{\url{https://cxc.harvard.edu/ciao/caveats/psf\_artifact.html}}. Such sub-arcsecond artifacts are the subject of ongoing calibration efforts and can be identified using the \texttt{make\_psf\_asymmetry\_region} script.}
\label{fig:psf_deconv}
\end{figure}

An example of a simulated on-axis point source and PSF-based image reconstruction is shown in Figure~\ref{fig:psf_deconv}, illustrating both the capabilities and limitations of these simulations on sub-arcsecond scales. While Figure~\ref{fig:psfsim} demonstrates the agreement for off-axis sources, Figure~\ref{fig:psf_deconv} shows the corresponding on-axis case, including PSF-based image reconstruction and the impact of sub-arcsecond instrumental features.

Together, ChaRT and MARX provide a flexible simulation environment that integrates naturally with CIAO workflows. By allowing realistic modeling of both the telescope optics and detector response, these tools play an important role in PSF analysis, instrument characterization, and the interpretation of complex X-ray observations. Their fidelity is, however, ultimately limited by uncertainties in the mirror geometry and its calibration, and by time-dependent effects such as contamination buildup on ACIS \citep{plucinsky2022}, which have become increasingly evident in deep observations, particularly in the sub-arcsecond structure of the PSF.

\sherpa\ supports spectral simulations through the \texttt{fake\_pha} command, which generates synthetic count spectra by folding a user-defined source model through instrument response files (ARF and RMF), accounting for exposure time, background components, and Poisson realizations of the expected counts. This capability enables realistic prediction of the Chandra ACIS spectrum for planning observations or for generating model inputs required for MARX or ChaRT simulations, and extends to other missions such as NuSTAR, allowing consistent spectral analysis and simulation across multiple instruments. Sherpa can also be used to simulate Poisson-distributed images by evaluating spatial models on a pixel grid and generating corresponding Poisson realizations, providing a way to test imaging analyses and model fitting in the low-count regime.

\section{Science Acceptance Testing}

Scientific testing of CIAO tools has been a core component of the development process since the early stages of the project \citep{karovska2006, lee2011}. Ensuring consistent and scientifically reliable results across platforms is essential for each CIAO release.

CIAO testing verifies both scientific correctness and software stability. A key requirement is that code revisions do not introduce regressions, and therefore a comprehensive test suite is maintained for all tools, supporting libraries, and contributed scripts. The regression test suite includes about 2000 individual tests, with results compared against platform-specific baselines to identify expected changes or unintended behavior.

\subsection{Unit Testing}

Unit tests are performed on individual tools or scripts to verify their scientific performance within realistic analysis paths. These tests evaluate the behavior of each component in the context of upstream and downstream processing steps, ensuring that outputs are scientifically meaningful and consistent when integrated into end-to-end analyses.

\subsection{Regression Testing}

Regression testing is used to ensure that updates do not change expected behavior. Test cases are based on validated unit tests and are run across all supported platforms. The process is automated, including running the tests, comparing results to baseline outputs, and flagging any differences within defined tolerances.

Testing proceeds iteratively: issues identified during regression testing are reported to developers, corrected, and re-tested until consistency is achieved. This process ensures stable and reproducible behavior across platforms and software releases.

\subsection{Release and Installation Testing}

In addition to scientific validation, CIAO releases undergo installation testing to verify the integrity of distributed builds. CIAO and the associated calibration database are installed on multiple platforms and configurations to ensure reliability and consistency for end users.

\section{Software Distribution and Installation}

CIAO is distributed as a complete, pre-built software package that includes the core analysis tools, the \sherpa\ modeling application, SAOImageDS9, contributed scripts, MARX, and the CALDB. The source code is available, and several components can also be obtained independently. This distribution model reflects the role of CIAO as a unified analysis system while preserving flexibility in how its components are accessed and used.

The installation model has evolved over time, from early manual distributions (via mailed compact discs) to more automated approaches, significantly simplifying setup for users \citep{fruscione2010,fruscione2014,fruscione2024}. Current releases are designed to provide a consistent and ready-to-use environment across supported platforms, minimizing the need for manual configuration. For advanced use cases, CIAO can also be compiled from source, providing additional flexibility for specialized environments.

CIAO is available for Linux and macOS systems, and its streamlined distribution has improved documentation for new users while maintaining the stability and reproducibility required for scientific analysis.

\section{Documentation and User Support}

Documentation and user support have played a key role in the widespread adoption of CIAO. From its inception, the system has emphasized accessibility, providing resources that allow users with a wide range of experience to perform complex X-ray data analysis.

CIAO documentation is organized as a structured set of complementary resources. Central to this arrangement are the Science Analysis Threads\footnote{\url{https://cxc.cfa.harvard.edu/ciao/threads/index.html}}, which provide task-oriented guides for common analysis workflows, including spectral extraction, source detection, timing analysis, and PSF modeling. These are complemented by the \textit{ahelp} system, which provides detailed reference documentation for individual tools, including parameter descriptions and usage examples \citep{burke2006,galle2003}. These components support both guided learning and user-defined analysis.

Additional documentation resources include high-level analysis guides\footnote{\url{https://cxc.cfa.harvard.edu/ciao/guides/index.html}} and specialized materials that explain the reasoning behind analysis choices\footnote{\url{https://cxc.cfa.harvard.edu/ciao/why/index.html}}. Dedicated web pages for key topics, such as PSF analysis\footnote{\url{https://cxc.cfa.harvard.edu/ciao/PSFs/psf\_central.html}} and multi-observation data merging\footnote{\url{https://cxc.cfa.harvard.edu/ciao/merging/merge\_central.html}}, further support users in addressing more complex tasks.

Beyond written documentation, user support is provided through the Chandra Helpdesk\footnote{\url{https://cxc.harvard.edu/help/}}, which serves as the primary interface between users and the CXC \citep{lee2024}. The Helpdesk handles several hundred requests per year, with a typical response time of a day, the majority related to CIAO and \sherpa\ data analysis. This interaction allows users to obtain guidance on both technical and scientific issues, while also providing feedback that informs ongoing development of the software and its documentation.

Documentation is further supported by training activities, including workshops\footnote{\url{https://cxc.harvard.edu/ciao/workshop/index.html}} that provide hands-on experience with real datasets. Online resources such as video tutorials\footnote{\url{https://www.youtube.com/user/4ciaodemos}} complement these efforts by providing introductions and updates on new features.

The documentation system is continuously refined based on user feedback and evolving analysis needs. This approach ensures that CIAO remains accessible while supporting increasingly sophisticated analysis techniques. As a result, documentation has played a key role in enabling a broad and active user community and in maintaining CIAO as a standard tool for X-ray data analysis.

\section{Using CIAO for Science}

A typical CIAO Chandra analysis session involves downloading data, reprocessing it with current calibration, performing field-of-view-wide observation analysis, and carrying out detailed studies of selected sources. As an illustrative example on how CIAO tools and applications work together, we outline the basic steps involved in the spectral analysis of an X-ray point source.

Chandra observations are identified by an integer Observation ID (ObsID). Relevant datasets can be located using CXC web interfaces (e.g., ChaSeR\footnote{\url{https://cda.cfa.harvard.edu/chaser/}}) or from the command-line using CIAO scripts such as \texttt{find\_chandra\_obsid}, which searches by source name, and \texttt{download\_chandra\_obsid}, which retrieves data for specified ObsIDs. Command-line tools provide access to public data, while proprietary observations must be obtained through authenticated web services.

Downloaded data are organized into directories named by ObsID, each containing \texttt{primary} and \texttt{secondary} subdirectories with the science data products. A standard first step is to reprocess the data using the \texttt{chandra\_repro} script, which applies the latest calibration and generates a \texttt{repro} subdirectory containing files ready for further analysis.

Initial data inspection is typically performed by creating an energy-filtered event file (e.g., 0.5--7 keV excluding lower and higher energy ranges which are usually background-dominated) using the \texttt{dmcopy} tool, and displaying it with the DS9 imaging application. Visualization tools in DS9, including scaling and color adjustments, are used to examine the field and identify sources. For more quantitative imaging, the \texttt{fluximage} script can be used to produce exposure-corrected images and flux maps, perhaps combined with adaptive smoothing tools to enhance low-surface-brightness features.

Source analysis proceeds by defining spatial extraction regions around interesting sources, often interactively in DS9 and saved as region files. Background regions are selected from nearby source-free areas. CIAO tools support region filtering through virtual file specifications (e.g., \texttt{evt.fits[sky=region(src.reg)]}) to isolate events within specified regions. The \texttt{specextract} script is then used to extract the source and background spectral count files along with the corresponding calibration products (ARF and RMF). Higher-level tools such as \texttt{srcflux} can additionally estimate the overall energy fluxes in physical units in a given energy band.

To interpret the spectral data, a model is fit to the instrumental spectrum using \sherpa, CIAO's modeling and fitting application. This process involves defining a parameterized physical model (for example, an emission model combined with interstellar absorption) and convolving it with the instrumental response to predict the observed signal. The model parameters are then iteratively adjusted to minimize a chosen fit statistic and obtain the best-fit solution. Within \sherpa, users load the extracted spectral dataset, select an appropriate model, and choose both the fit statistic and optimization method. The \texttt{fit} command is used to determine the best-fit parameters, while routines such as \texttt{reg\_proj} can be applied to estimate confidence intervals. Derived quantities, including integrated energy fluxes over specified bands, can be computed using tools such as \texttt{calc\_energy\_flux}. These analysis steps can also be carried out through the DAX interface, which provides a CIAO graphical environment within DS9 for source analysis, including region selection, spectral extraction, model definition, fitting, and visualization of results, as illustrated in Figure~\ref{fig:dax}. Both \sherpa\ and DS9 also provide capabilities for generating publication-quality figures to present the results.

\section{Current Limitations and the Future Ahead}

CIAO development began almost thirty years ago as a mission-specific system for Chandra, and its architecture still carries the imprint of that origin. Many of its core design choices---local installations, command-line interfaces, reliance on local parameter files and tight integration with mission-specific calibration---would likely be reconsidered if the system were developed today. For example, while the command-line paradigm provides flexibility and can be scripted, it can present a significant barrier to entry for new users and remains less accessible than more interactive, notebook-based analysis environments. Despite this, CIAO has continued to evolve, incorporating new tools, scripting environments, and extensibility while maintaining continuity for its user community. It remains a widely used and reliable analysis environment, a testament to both its original design and development over time. Looking forward, future development may include transition from local to distributed analysis, including cloud-based platforms such as SciServer \citep{popp2020} or Fornax\footnote{\url{https://science.nasa.gov/astrophysics/programs/physics-of-the-cosmos/community/the-fornax-initiative/}} and remote, browser-accessible visualization through SAOImageDS9.

At the same time, new standards such as the Model Context Protocol (MCP), which enable AI models to interact with external data, tools, and software systems, may offer new ways of interacting with CIAO, potentially exposing it to a wider community and allowing its integration into large-scale, multi-wavelength data analysis procedures.

\section{Recommended Software Citations}
\label{sec:cite}

Users of CIAO or any of the subsystems described in this paper are encouraged to cite the following references in publications based on their work. If space is limited, citing the most recent reference is sufficient:

\begin{itemize}[label={}, labelsep=0pt, leftmargin=2pt, itemsep=1pt, topsep=2pt]
\item CIAO: this paper and \citealt{fruscione2006}
\item Sherpa: \citealt{siemiginowska2024}
\item DS9: \citealt{fruscione2026} and \citealt{joye2003}
\item MARX: \citealt{davis2012}
\item ChaRT: \citealt{carter2003}
\item SAOTrace: \citealt{jerius2004}
\end{itemize}

Additionally, authors are encouraged to report the CIAO, CALDB, and relevant subsystem versions used in their analysis.

\begin{acknowledgments}
CIAO would not be what it is today without the contributions of numerous members from the CXC Science Data System (SDS) group and Data Systems (DS) division. Their collective efforts have been central to the development and evolution of the system over the years.

We acknowledge the contributions of past members of the SDS group at the Center for Astrophysics $|$ Harvard \& Smithsonian (CfA), listed here in alphabetical order: Nina Bonaventura, Nancy Brickhouse, Catherine Cranmer, Nick Durham, Adam Dobrzicki, Martin Elvis, Elizabeth Galle, Dan Harris, Holly Jessop, Margarita Karovska, Casey Law, Fabrizio Nicastro, Frank Primini, Jennifer Rittenhouse West, Eric Schlegel, Randall Smith, and Olaf Vancura. We also recognize past contributors from the Massachusetts Institute of Technology (MIT): Glenn Allen, John Davis, Dave Davis, John Houck, Joel Kastner, Bish Ishibashi, Mike Noble, Mike Nowak, and Michael Wise.

We further acknowledge the work and expertise of the current and past members of the CXC Data Systems division. In particular, we recognize the contributions of the current members (in alphabetical order): Mark Chaoui, Harlan Cheer, Giuseppina Fabbiano (Division Head), Danny G. Gibbs II, Dale Graessle (Chandra CALDB Manager), Helen He, Omar Laurino, Janine Lyn, and Charles Paxson, as well as the many dedicated contributors to the division over the past 26 years, including Stephen Doe, Dan Nguyen, James Overly, Brian L. Refsdal, Christopher Stawarz, and Marie Terrell. We are deeply grateful for the dedication and expertise of all these individuals, whose collective efforts have made CIAO an invaluable resource for the X-ray astronomy community.

We also acknowledge the CHASC International Astrostatistics Center (CHASC), an interdisciplinary collaboration that has played an important role in advancing statistical methods for high-energy astrophysical data analysis. In particular, we thank David Van Dyk, Vinay Kashyap, Xiao-Li Meng, and Thomas Lee for their sustained engagement and contributions to this effort.

The authors thank the anonymous referee for carefully reading the manuscript and for providing insightful comments.

This work was supported by NASA contract NAS8-03060 to the CXC, which is operated by SAO for and on behalf of NASA. HMG, DH, MN and DAP were supported by contract SV3-73016 to MIT for support of the CXC.

This research has made extensive use of the Astrophysics Data System, funded by NASA under Cooperative Agreement 80NSSC21M00561.

Generative AI tools were used during the preparation of this manuscript for editing, LaTeX formatting, and consistency checks. The authors maintain full responsibility for the scientific content and the accuracy of the final text.
\end{acknowledgments}

\begin{contribution}
The development, evolution, and maintenance of the CIAO system represent the collective work of the CXC Science Data System group and Data Systems division over more than 25 years. While all authors and team members listed below contributed significantly to the project's long-term success, specific roles for this manuscript are as follows: \\
A. Fruscione served as the primary author and led the preparation and submission of the manuscript.\\
K. Glotfelty authored the technical Appendix.\\
A. Siemiginowska, D. Huenemoerder, H. M. G\"unther, J. McDowell, K. Glotfelty, N.P. Lee provided the primary technical feedback, critical revisions, and oversight for the paper.\\
D.A. Principe, D. Burke, M. Cresitello-Dittmar, I.N. Evans, J.D. Evans, W. Joye, W. McLaughlin, J.B. Miller, and M. Nynka contributed to the ongoing development, testing, and documentation of the CIAO components described in this work.
\end{contribution}

\vspace{5mm}
\facilities{Chandra X-Ray Observatory, XMM-Newton, NuSTAR, SOHO, Hubble, Spitzer, SDSS, Gemini}
\software{CIAO, Sherpa, SAOImageDS9, MARX, ChaRT, SAOTrace, Matplotlib, Python, NumPy, conda, HEASoft, XSPEC, SAS, FTOOLS, IRAF}

\appendix

\section{The Internals of CIAO}

CIAO is engineered with a modular and robust architecture, composed of a subset of the same tools used in the CXC standard data processing pipelines. The tools are portable, open source, and regularly updated to meet evolving user system requirements and preferences. The primary implementation languages are C and C++. The CIAO core tools also include a small number of Bourne shell and Python scripts, while the CIAO contributed scripts package consists almost entirely of Python scripts. With over 200 compiled tools and scripts developed over more than 25 years by dozens of developers, consistency is important. CIAO achieves this by using several core libraries.

CIAO uses a familiar command-line user interface similar to its contemporaries. Like FTOOLS and IRAF, parameter and value pairs are stored in text parameter files that can be accessed and manipulated with utility tools. The CXC parameter interface library (\texttt{cxcparam}) provides a common interface to all compiled tools and Python scripts to interact with parameter files. CIAO differs from IRAF in the absence of a host application (e.g. the \texttt{cl} environment); instead, CIAO tools are run from the user's shell of choice, making it easier to put into pipelines and scripts and to integrate with other non-CIAO tools.

Data processing is file-based, centering on event files, which are FITS binary tables where each row provides information about a single event (e.g., a single photon) such as its position, energy, time, and some metrics of quality (grade, status, etc.). The CXC Data Model (DM) library provides an abstraction layer to access FITS and ASCII files without having to know the details of the underlying format. For example, instead of having to know to read the \texttt{NAXIS2} keyword to get the number of rows in a FITS table, the CXC Data Model provides a generic routine: \texttt{dmTableGetNoRows}. Earlier versions of the DM also supported the IRAF ``QPOE'' format used by legacy missions; this is no longer required, as these data are now available in FITS format via HEASARC\footnote{\url{https://heasarc.gsfc.nasa.gov/}}.

Chandra data processing is heavily dependent on numerous calibration products for everything from instrument geometry, plate scale, effective area, vignetting, spectral response, point spread function, astrometric alignment, and much more. The CIAO CALDB library (\texttt{caldb4}) extends the standard HEASARC Calibration Database (CALDB) definition and interface to accommodate Chandra's processing requirements. The Chandra CALDB is used to automatically identify and locate the applicable calibration files needed for a specific observation based on the observation date and additional criteria such as focal-plane temperature, operating mode, etc.

Significant emphasis is placed on preserving processing provenance and managing metadata. CIAO tools record the full set of parameter values in \texttt{HISTORY} records, which can then be easily read with the \texttt{dmhistory} tool to recreate the exact commands used to create a file. \texttt{HISTORY} records are retained from input file to output file. By using a common header library, CIAO tools also ensure that a sufficient set of keywords are propagated from input to output and are updated as appropriate (e.g., when combining multiple input files into a single output file). The DM subspace is maintained in a file's metadata, which is also retained.

Additional libraries provide single implementations for common algorithms and calibrations. The pixel library (\texttt{pixlib}) provides routines to perform all coordinate transformations internal to Chandra. This includes calibrations of the telescope configuration (mirror shells, focal length, etc.), Science Instrument Module (SIM) translation table, and detector geometry. All calibration calculations involving the Chandra effective area are handled by the Analysis Reference Database library (\texttt{ardlib}), which provides routines to determine mirror effective area and vignetting and detector efficiency as a function of location, energy, and time. The CXC \texttt{region} library contains routines to operate on regions, which include reading and writing FITS and ASCII (e.g., DS9 format) region files, region intersections and unions, calculating area, and determining if a point is included in the region. Consistency in error reporting and warnings, both content and format, is achieved by using the CXC \texttt{error} library. Finally, parameters that take multiple inputs use the \texttt{stack} library which supports wildcards (e.g., \texttt{*.fits}), list files, grids, and more. With these libraries, CIAO users enjoy a uniform and predictable user experience.

Chandra data are intrinsically heterogeneous, which means that CIAO tools need to be especially robust to a wide range of datasets. A single Chandra observation may run from 1 ks up to 160 ks. Data may need to be combined across a 25-year baseline. Targets range from nearby moving targets (Earth, moon, comets, planets) to stars, galaxies, clusters, and beyond. Observers can choose one of two science instruments, possibly with one of two gratings inserted, with numerous operating modes (different CCDs, SIM offsets, TIMED vs.\ Continuous Clocking, Subarray, Window, Interleaved, etc). Most targets are photon-starved, requiring Poisson statistics for analysis; those bright enough to assume Gaussian statistics may suffer from pileup. CIAO tools have rigorous requirements to ensure that they are scientifically valid under a wide range of conditions.

CIAO tools adhere to various Interface Control Documents (ICDs) which prescribe the required input formats and expected output formats for all files. CIAO tools also follow a common lexicon for all standard metadata (header keywords) that specifies their origin, meaning, units, applicability, and grouping\footnote{\url{https://cxc.cfa.harvard.edu/ciao/data_products_guide/ascfits.pdf}}. Tools also adhere to a common schema that maps data products to unique database identifiers\footnote{\url{https://cxc.cfa.harvard.edu/contrib/arots/fits/content.txt}}, which enable search-and-retrieve interfaces to access Chandra's datasets with single-file granularity.

By adhering to community standards\footnote{\url{https://heasarc.gsfc.nasa.gov/docs/heasarc/ofwg/ofwg_recomm.html}}, CIAO tools are highly interoperable with data from other missions and generic data sources. This includes the event-list format\footnote{\url{https://heasarc.gsfc.nasa.gov/docs/heasarc/ofwg/docs/events/ogip_94_003/ogip_94_003.html}}, spectrum files (PHA\footnote{\url{https://heasarc.gsfc.nasa.gov/docs/heasarc/ofwg/docs/summary/ogip_92_007_summary.html}}), and spectral response files (ARF and RMF\footnote{\url{https://heasarc.gsfc.nasa.gov/docs/heasarc/caldb/docs/summary/cal_gen_92_002_summary.html}}). In situations where no standard existed, the CXC proposed new recommendations such as the FITS Region format\footnote{\url{https://fits.gsfc.nasa.gov/registry/region.html}}. Interoperability is maintained when possible; for example, FTOOLs can list the contents of a CIAO-created file. However, because FTOOLS does not store subspace information, CIAO tools cannot determine how a file has been filtered by FTOOLS \texttt{ftcopy}, which can affect calculations such as region area calculations.

With this solid foundation, CIAO has proven to be a robust software system, and continues to support the evolving Chandra mission and meet the needs of more modern systems.

\newpage
\bibliography{ciaobibliography}{}
\bibliographystyle{aasjournalv7}

\end{document}